\documentclass[12pt, preprint]{aastex}
\usepackage{multicol, color}

\begin{document}
\title{Dark Matter Halo Mergers I: Dependence on Environment  \& Redshift Evolution} 

\author{J. A. Hester\altaffilmark{1}, A. Tasitsiomi\altaffilmark{2}}
\email{jhester@srl.caltech.edu}

\altaffiltext{1}{California Institute of Technology, Pasadena, CA 91125}

\altaffiltext{2}{Lyman Spitzer Jr. Fellow, Department of Astrophysical Sciences, Peyton Hall, Princeton University, Princeton, New Jersey, 08540}

\begin{abstract}
This paper presents a study of the specific merger rate as a function of group membership, local environment, and redshift in a very large, $500h^{-1} Mpc$, cosmological N-body simulation, the \textit{Millennium Simulation}. The goal is to provide environmental diagnostics of major merger populations in order to test simulations against observations and provide further constraints on major merger driven galaxy evolution scenarios. A halo sample is defined using the maximum circular velocity, which is both well defined for subhalos and closely correlated with galaxy luminosity. Subhalos, including the precursors of major mergers, are severely tidally stripped.  Major mergers between subhalos are therefore extremely rare.  Tidal stripping also suppresses dynamical friction, resulting in long major merger time scales when the more massive halo does not host other subhalos.  In contrast, when other subhalos are present major merger time scales are several times shorter.  This enhancement is likely due to inelastic unbound collisions between subhalos. Following these results, we predict that major mergers in group environments are dominated by mergers involving the central galaxy, that the specific merger rate is suppressed in groups, and that the frequency of fainter companions is enhanced for mergers and their remnants.  We also observe an `assembly bias' in the major merger rate in that mergers of galaxy-like halos are slightly suppressed in overdense environments while mergers of group-like halos are slightly enhanced.  A dynamical explanation for this trend is advanced which calls on both tidal effects and interactions between bound halos beyond the virial radii of locally dynamically dominant halos.
\end{abstract}


\section{Introduction}
The ongoing mergers of both dark matter halos and the galaxies they contain is an inevitable component of hierarchical structure formation.  The potential impacts of these mergers on galaxy evolution are both varied and heavily debated. Mergers can be roughly divided into two classes; minor mergers, in which a small halo is accreted by a substantially larger halo, and major mergers, in which the two halos are of roughly similar mass.  The division between the two is usually placed near a mass ratio of 3:1.  Minor mergers contribute both stars and gas to forming galaxies, and are important for understanding the detailed morphologies of galaxies, particularly spiral galaxies.  Tidal forces during a minor merger may heat the thin stellar disk and drive bar instabilities, thick disks may also represent the remnants of disrupted satellites~\citep{Steinmetz02, Yoachim08}.  The effects of a major merger are likely more dramatic.  Major mergers between two gas rich spiral galaxies are a popular mechanism for creating elliptical galaxies, an idea dating back to~\citet{Toomre72}.  More recently, they have been invoked as a means of fueling intense starbursts and luminous AGN~\citep[][among others]{Mihos96, DiMatteo05, Springel05b, Springel05c}.

Theories connecting major mergers, starbursts, AGN, and the creation of elliptical galaxies have not been suitably tested observationally.  Theoretical models can reproduce the luminosity function and number density evolution of AGN while correctly recovering correlations between black hole masses and spheroid dynamics~\citep{Kauffmann00, Wyithe02, Wyithe03, Hopkins06}.  These models include assumptions about AGN and starburst lifetimes and efficiencies which tend to be under constrained by the observational data.  Additional constraints on these models are needed, as are model independent tests of major merger driven evolutionary scenarios. Environment is a potentially powerful probe which has been under exploited in previous work on major mergers.  To make use of environment, the environmental dependences of the major merger rate must be well understood.  This work therefore explores the environments of major mergers in the \textit{Millennium Simulation}, a large N-body simulation with a box length of $500h^{-1}\rm{Mpc}$ and a particle mass below $10^9M_{\odot}$~\citep{Springel05ms}.  This simulation is large enough to probe the full range of environments and has a fine enough mass resolution to follow galaxy-like halos and subhalos.  The goal of this project is to provide the theoretical groundwork necessary to use environment as a probe of merger driven galaxy evolution scenarios.

\subsection{Major Mergers, AGN, \& Morphology}
\label{merger scenario}

The link between major mergers, starburst and AGN fueling, and galaxy morphology has been extensively studied using simulations, and there is strong circumstantial observational evidence that these populations are correlated.  Simulations of individual major mergers involve imbedding a stellar and gaseous disk into each of two dark matter halos, placing them on a collision course, and observing the stages of the simulated merger.  Such models vary considerably in complexity, ranging from including only stellar disks to modeling multiphase gas disks.  They also vary in prescriptions for AGN and star formation fueling. These simulations capture the physics behind the major merger scenario and can predict the appearance of the different stages of a major merger~\citep[for examples see][]{Mihos96, Springel00, Naab03, Barnes04, Springel05c, Cox06}.  Complimentary simulations invoke semi-analytic prescriptions to introduce simplified gas physics into large dark matter simulations.  These allow the entire population of AGN, starbursts, and galaxy morphologies to be studied~\citep{Kauffmann00, Wyithe02, Wyithe03}.  

The results of previous theoretical studies of major mergers can be briefly summarized as follows.  The tightly bound stellar components remain in the center of their dark matter halos,  loose angular momentum through gravitational interactions with their own dark matter halos, and remain at the center of the merger remnant.  The two stellar components encounter each other with a tightly bound orbit, quickly merging after the cores of the dark matter halos have merged. The resulting stellar remnant is pressure supported rather than rotationally supported.  When no gas component is included, the central densities of simulated major merger remnants are lower than the observed cores of elliptical galaxies.  When gas is included, tidal forces funnel the gas into the centers of the merging galaxies prior to the galaxies merging.  This gas then settles to the center of the merger remnant.   Of the initial gas disk, $\geq 50\%$ ends up in the central component of the merger remnant~\citep[see review in][]{Barnes92}.  Allowing this gas to form stars builds a stellar core in the remnant, similar to that seen in elliptical galaxies~\citep{Mihos96}.  Gas that is funneled to the centers of the merging galaxies can also fuel AGN activity both in the merging galaxies and in the remnant.  When black hole growth and star formation in the merger remnants both occur with efficiencies that are proportional to the gas fraction in the central region, semi-analytic models reproduce observed correlations between black hole masses and spheroid masses~\citep{Kauffmann00}.  Self regulated models of black hole growth go a step further. With abundant fuel, the black hole at the center of the remnant grows exponentially both in mass and in luminosity.  Feedback from the growing AGN heats and drives winds in the surrounding gas and can become strong enough to expel gas from the center of the remnant ~\citep{Silk98, Fabian99}, terminating both star formation and AGN activity.  Self regulated models of black hole growth predict that the ultimate mass of the black hole is strongly correlated with the depth of the central gravitational potential~\citep{DiMatteo05, Springel05b, Springel05c}.  Furthermore, nuclear star formation and black hole growth are both regulated by the same process~\citep[e.g..][]{Hopkins06}.  The resulting major merger remnant has a stellar profile which resembles an elliptical galaxy, hosts a central super massive black hole with a mass that correlates both with the central potential and the central stellar density, is gas poor, and has a diffuse halo of hot gas.  If this gaseous halo is prevented from cooling and forming stars, the post-starburst remnant will fade to become `red and dead'.  Feedback between cooling in the nucleus and low level AGN activity is often invoked to prevent further star formation in the remnant~\citep{Best06, Fabian06}.

There is observational evidence to support this merger driven evolutionary scenario.  Ultraluminous infrared galaxies (ULIRGS) represent powerful starbursts.  Observations of ULIRG morphologies indicate that they are ongoing or recent major mergers~\citep[see reviews in][]{Sanders96, Jogee04}.  Radio observations of these galaxies indicate that they have central concentrations of dense cool gas~\citep{Scoville86, Sargent87, Sargent89}.  Powerful IR selected starbursts are also known to host obscured or low luminosity AGN~\citep{Komossa03, Gerssen04, Alexander05, Borys05}. The physical number density of luminous AGN rises with redshift, peaking around $z\sim2-3$~\citep{Boyle00, Fan01}, similar to the major merger rate of massive halos in N-body simulations.  Black hole growth and spheroidal growth appear to be closely related.  Black hole masses correlate both with the mass and the velocity dispersion of the spheroids, either elliptical galaxies or spiral bulges, that host them~\citep{Magorrian98, Ferrarese00, Gebhardt00, McLure02, Tremaine02, Marconi03}.  Post-starburst E+A (or K+A) galaxies have high central surface densities, kinematically hot older stellar populations, and will fade to resemble elliptical galaxies in the absence of additional star formation~\citep{Norton01, Yang04, Goto05}.  Post-starburst galaxies also frequently display disturbed morphologies, indicative of mergers or tidal interactions~\citep{Yang04, Goto05, Owers07}.  Elliptical galaxies are found preferentially in group and cluster environments~\citep{Dressler80, Norberg02, Hogg03, Kauffmann04, Blanton05b}, and simulated dark matter halos in equivalent environments preferentially experienced major mergers in their pasts~\citep{GKK}.

We intend to use a dark matter simulation to develop environmental diagnostics of major merger populations that can be used to test the merger driven galaxy evolution scenario.  By doing so we are assuming both a one to one correspondence between dark matter mergers and galaxy mergers and that the relevant dynamics are dominated by the dark matter. As discussed above, when the cores of two dark matter halos that each host a galaxy merge, the galaxy merger is immanent.  The final stages of the galaxy merger occur quickly; \citet{Cox06} find that the final galaxy merger in a 1:1 merger takes $\approx 200$Myr.  While there has been some discussion of `dark halos' which do not host galaxies~\citep[][and references therein]{Maccio06}, this occurs at $V_{\rm max}$ well below those considered here.  Similarly, their is no observational evidence for orphan galaxies~\citep{Mandelbaum06}, and truncated dark matter halos have been observed around galaxies in clusters~\citep{Natarajan07}.  A one to one correspondence between halo mergers and galaxy mergers is therefore a reasonable assumption.   While the inclusion of baryons might affect some of the relevant dynamics, dark matter constitutes a strong majority of the matter.  Hence, the dynamics is dominated by the dark matter, with possible refinements to be introduced by including baryons.  Exploring this issue is a potential topic for future work.

\subsection{Mergers \& Environment - Previous Results}

The earliest, and simplest, theoretical studies of the merger rate were based on extended Press Schechter theory, which was in turn based on linear theory plus spherical collapse models~\citep{Press74, Bond91, Lacey93}. Mergers between halos were assumed to occur on the time scale of dynamical friction and mergers between subhalos were neglected.  With these assumptions, merger trees can be built analytically for all halos existing today~\citep{Kauffmann93, Somerville99}.  The assumption that mergers occur on a dynamical friction timescale can also be applied to N-body simulations when subhalos cannot be resolved.  Under this assumption,~\citet{Kauffmann00} find that the major merger rate is independent of local environment.  Extensions of this treatment first considered the merger rate between subhalos, concluding that the sub-sub merger rate could be quite high within group mass hosts.  Further work began to include dynamical effects within the host halo such as tidal stripping.  This was done both analytically and with N-body simulations~\citep{Mamon00, Penarrubia05, Boylan07}.  Assuming that linear theory is modified only by dynamics within virialized halos, that is that linear theory correctly describes accretion histories, all non-linear effects are confined to within host halos. In this treatment, correlations between the merger rate and environment on scales beyond the virial radii of the hosts, measured either directly or through clustering, arise from convolving any correlations between merger rate and host mass with the clustering of the hosts.  In general, linear theory breaks down when tidal forces become important, so some deviations should be expected.  

Previous treatments of the merger rate between subhalos have found that the specific merger rate, that is the merger rate per halo or galaxy, should be enhanced in groups and suppressed in clusters.   Within a bound virialized halo, where it may be possible to assume relative subhalo velocities are random, the specific major merger rate between subhalos can be expressed as
\begin{displaymath}
R_m = n_h\langle \sigma_m v \rangle
\end{displaymath}
where $\sigma_m$ is the merger cross section, $v$ is the relative velocity of the halos, and $n_h$ is the number density of potential interaction or merger partners.  Generically,  the merger cross section must increase with subhalo mass and be highly dependent on the relative velocities of the subhalos.  In clusters, where relative subhalo velocities substantially exceed the internal velocities of the subhalos, merger rates between subhalos are likely suppressed relative to distinct, or non-sub, halos.  Groups however appear to represent an environment which combines higher than average halo densities with relative velocities that are conducive to merging.  

Merger cross sections based on simulated encounters between distinct halos of equal mass peak near the internal velocity of the halos~\citep{Makino97}.  Using these cross sections and assuming a Maxwellian velocity distribution for the subhalos that scales appropriately with host halo mass, \citet{Makino97} find that merger rates are enhanced in groups and decline with the mass of the host halo, becoming suppressed in clusters.  In this treatment the subhalo merger rate also increases with the mass of the subhalos, as the merger cross section increases.  Analytical and numerical treatments indicate that tidal stripping occurs on a much shorter time scale than dynamical friction, resulting in a population of tidally stripped subhalos for which dynamical friction is ineffective ~\citep{Mamon00, Penarrubia05, Boylan07}. In a mass defined subhalo sample, tidal stripping reduces the subhalo number densities by removing subhalos from the sample.  In a subhalo sample based on pre-accretion masses, tidal stripping can drastically reduce the subhalo merger cross section. Previous work has taken the first approach. When subhalo populations are selected using post tidal stripping masses, the specific merger rate is extremely sensitive to the host halo mass and is still enhanced in groups and suppressed in clusters~\citep{Mamon00}.   

Numerical simulations support these analytical results.  \citet{Ghinga98} simulate a large cluster with a high resolution in order to observe subhalos.  They find that halos cease merging once they enter the cluster. \citet{DeLucia04} also find that the merger rate drops after sub-halos are accreted by a cluster.  \citet{GKK} make a complimentary measurement in a cosmological N-body simulation.  They measure the merger rate of the most massive progenitors of the halos identified at $z=0$, therefore not counting major mergers that result in remnants that later merge with a more massive halo.  They find that halos that reside in clusters at $z=0$ have the lowest merger rates near $z=0$, but had merger rates higher than the progenitors of isolated halos in the past.  The merger rates for halos that reside in groups at $z=0$ are higher than for isolated galaxies. 

Studying the merger rate as a function of host halo mass is attractive because it has direct observational consequences.  Doing so may also capture most of the non-linear physics affecting the merger rate.  Under the assumptions of linear theory, once a halo mass is specified, in this case the mass of the host halo, the accretion history is independent of environment~\citep{White96}.  Several recent studies of halo properties have indicated that accretion history does have a residual dependence on local environment~\citep{Wechsler06, Croton07, Gao07}.  There may be a similar dependence between the major merger rate and local environment.  This has not yet been studied.  

While the environments of major mergers are clearly a widely studied topic, the \textit{Millennium Simulation} should allow us to make an important advancement. The \textit{Millennium Simulation}, with its superb combination of size and resolution, allows the study of all of the above issues in concert.  We will use a common, well defined, language to study these issues and, in addition, should be able to see the effects of the interplay between them.  Finally, we will focus not only on studying  the dynamics of major mergers, but on using definitions of environment and the major merger rate that have clear observational counterparts.   This is essential as the ultimate goal is to use the results of this work to craft observational tests that are capable of identifying merger populations.

\section{Methods}
\label{methods}
\subsection{Simulation \& Numerical Issues}
\label{description of simulation}
Our study is performed using the {\em Millennium Simulation} (MS) \citep{Springel05ms} run using a version of GADGET2 \citep{Springel05g}. The MS is a cosmological N-body simulation of the $\Lambda$CDM universe that follows the evolution of more than 10 billion particles in a box of 500$h^{-1}$ Mpc comoving on a side. The particle mass is $8.6\times10^{8} h^{-1} M_{\odot}$, and particle-particle gravitational  interactions are softened on scales smaller than 5$h^{-1}$kpc.  The simulation uses  parameters in agreement with the WMAP1 results \citep{Spergel03}: $\Omega_{m}=0.25$,  $\Omega_{\Lambda}=0.75$, $h=0.73, n=1$ and $\sigma_{8}=0.9$. 

We use halos drawn from the MS halo catalog.  The first step in halo identification is a friends-of-friends (FOF) group finder which is built into the simulation code.  Particles separated by less than 0.2 times the mean particle separation are grouped together in a FOF group.  This combines particles into groups with a mean over-density which is somewhat lower than the expected overdensity of virialized halos at low z and approaches the expected overdensity as z increases.   In post-processing, bound halos are identified within these FOF groups using an improved version of the SUBFIND algorithm \citep{Springel01}.   In each FOF group both a dominant central halo and its subhalos are identified. SUBFIND first computes the smoothed density field at the positions of all the particles.  It then defines as possible halos all regions centered around locally overdense points that are bounded by the first isodensity contour to traverse a saddle point in the density.  Each halo candidate is subjected to a gravitational unbinding procedure.  Structures that retain greater than 20 particles are kept in the halo catalog and their basic properties are determined. 

In this study we characterize halos by the maximum of their rotation velocity curve, $V_{\rm max}$, rather than by their mass.  This is largely motivated by our interest in tracking subhalos. Defining mass in the case of subhalos can be problematic.  It is not clear that SUBFIND provides reliable subhalo masses, see discussion in~\citet{Natarajan07}, and, regardless of the specifics of the halo finder, subhalo mass is itself a relatively ill-defined concept.   In addition, we are interested in comparing the simulation to observations and are therefore concerned primarily with the hypothetical luminosity of the galaxy hosted by the halo or subhalo, and the `correct' halo measure is therefore the quantity that is most closely correlated with this luminosity. Even a well defined `tidal mass' is therefore not a suitable measure.  When subhalos are accreted they undergo substantial tidal stripping, but in the majority of cases the central galaxy remains intact while the surrounding dark matter halo decreases with mass.  The initial correlation between halo mass and galaxy luminosity therefore no longer applies to subhalos.

Using $V_{\rm max}$ as a proxy for galaxy luminosity is well motivated and does not suffer as strongly from the issues mentioned above.  The use of $V_{\rm max}$ is justified by an array of successful comparisons between observations and collision-less simulations in which halos are assigned a galaxy luminosity by associating luminosity and $V_{\rm max}$ \citep[e.g.,][]{Kravtsov04a,Tasitsiomi04,Tasitsiomi08}. These studies make a case for the hypothesis that, as a measure of the central halo potential, $V_{\rm max}$ quantifies the ability of a halo to allow baryons to cool and form stars.  Note that while $V_{\rm max}$ may not be the only such quantity, these studies indicate that it is the {\em dominant} quantity.  As a measure of the central halo potential, $V_{\rm max}$ is also significantly better defined for subhalos, thus overcoming a technical issue associated with using tidal masses.  While $V_{\rm max}$ itself is not immune to the affects of tidal stripping, it has been shown to responds to tidal stripping much slower than mass \citep[e.g.,][]{Kravtsov04b}.
 
The results presented here are sensitive to the completion of the halo catalog, which includes both distinct halos and subhalos. The halo catalog with which we work includes halos with $V_{\rm max}\ge120\rm{km/s}$, though our mergers may have progenitor halos with $V_{\rm max}$ down to $100\rm{km/s}$. The larger complete halo catalog from the Millennium Simulation includes all halos with greater than 20 particles, or a bound mass greater than $1.7\times10^{10}M_{\odot}$.  We test for a flattening of the $V_{\rm max}$ function in our halo catalog and find no sign of incompletion down to $V_{\rm max}=120\rm{km/s}$ at $z=0$ or $z=4$.  The relationship between $V_{\rm max}$ and mass for distinct halos evolves with redshift in that as redshift increases $M_v(V_{\rm max})$ decreases.  Mass scales approximately like $V_{\rm max}^3$ at all redshifts, as expected.  At redshift $z=0$, we expect an average distinct halo with $V_{\rm max}=100\rm{km/s}$ to have a virial mass $M_v\approx1.2\times10^{11}M_{\odot}$ and include $n_p\approx150$ bound particles.  At $z=4$, an average distinct halo with $V_{\rm max}=100\rm{km/s}$ has a virial mass $M_v\approx3.3\times10^{10}M_{\odot}$ and includes $n_p\approx38$ bound particles.   Given the relatively large average particle number, we can be reasonably sure that at $z=0$ distinct halos are complete down to  $V_{\rm max}=100\rm{km/s}$.  At $z=4$ completion of even distinct halos may begin to be an issue. In general, the effects of halo completion on our results are redshift dependent.

While the particle mass resolution of the MS is fine enough for distinct halos, subhalos above the smallest $V_{\rm max}$ of interest may suffer from over-merging if inadequate mass and force resolution allow subhalos to be prematurely disrupted.  As a subhalo is stripped and the number of bound particles decreases, two-body interactions can cause a subhalo to artificially evaporate, and hence the mass resolution can limit the mass, or $V_{\rm max}$, to which subhalos can be reliably identified.  Similarly, inadequate force resolution results in subhalos with artificially large cores which are easily disrupted and artificially lost.  These problems can be solved with increased mass and force resolution.  Ideally one would rerun the MS with different mass and spatial resolution to estimate the impact of these effects, but this route is prohibitively expensive.  These issues have initiated multiple numerical studies, and as an alternative to rerunning the simulation we will rely on the results of~\citet{Klypin99}.  Their results agree with other, similar, numerical studies. \citet{Klypin99} provide a pair of criteria that must be fulfilled in order for over-merging of subhalos not to be important.  The authors motivate their conditions analytically before verifying them with ART simulations.  In principle they are therefore independent of the specifics of the simulation.  They find that at least 20-30 particles are required in order for two-body evaporation to be negligible and that the tidal radius of the subhalo must be at least as large as a couple of spatial resolution lengths.  Note that the 20-30 particles refers to the number of bound particles post-tidal stripping.  These limits are not exact, but they highlight the relevant concerns and will be used to indicate approximately when subhalo completion becomes an issue. The possible effects of subhalo completion on our individual results are discussed in detail in Section~\ref{completion}.

The final simulation issue which must be considered is the spacing of the simulation outputs for which the halo finder is run.  There are 64 saved outputs, most of which are equally spaced in $\log(1+z)$  between z=20 and z=0. The typical time elapsed between two consecutive outputs is 2-400 Myr.  Typical time scales for an accreted halo to merger with its host halo are a few to 10 Gyr~\citep{Boylan07}, though we find that in some cases this time scale may be shortened by approximately a factor of 10.  Therefore, except in some extreme cases, accreted halos are resolved as subhalos in multiple outputs before a merger with the host occurs.  This temporal resolution distinguishes this work from some past studies \citep[e.g.,][]{Klypin99}.

\subsection{Merger Trees}

Every halo in the MS is associated with a merger tree that contains all of the progenitors of that halo during each previous simulation output.  The merger tree therefore contains the merger history of each halo.  Mergers trees are constructed after halos are identified in the 64 saved outputs.  The spacing of these outputs is dense enough to reliably identify the descendant of nearly every halo in the following timestep. Halos are required to have a unique descendant in either the following timestep or, in a small number of cases, the time step after. Identifying all descendants at each timestep defines the merger trees.  Descendants are identified by tracking particles, each of which is labeled.  Every halo in the next timestep that shares particles with the halo under consideration is identified.  These particles are weighted according to how bound they are to the halo under consideration, with tightly bound particles receiving a higher weight.  The halo in the following timestep with the highest weighted sum is marked as the unique descendant.  This scheme aims at tracking the inner cores of the halos which are less vulnerable to mergers and tidal stripping.  The construction of the merger trees is described in the supplementary information to~\citet{Springel05ms}.

The merger trees are used to identify major mergers.  While all halos have a unique descendant, when a merger has occurred halos have two or more progenitors in the previous timestep.  For this work, major mergers are defined using the $V_{\rm max}$ ratio of two merger progenitors.  The progenitor with the higher $V_{\rm max}$ is referred to as the MMP, or most massive progenitor, and that with the lower $V_{\rm max}$ as the LMP, or least massive progenitor.  For halos that are not the subhalo of a larger halo, $V_{\rm max}$ correlates closely with mass.  Major mergers are defined as having $V_{\rm LMP}/V_{\rm MMP}>0.7$.   In the case of halos that are not subhalos, this is the equivalent of mergers with less than a 3:1 mass ratio.  In the case of subhalos, the mass ratio of the halos may vary considerably, but the luminosity ratio of the galaxies typically hosted in the halos will not.  

Two definitions of a specific merger rate are used in this paper, where a specific merger rate measures the number of mergers per Gyr normalized by the number of halos, as opposed to a physical number density of mergers per Gyr.  The first rate is a backwards looking merger rate, $R_-$ which is defined as the number of major mergers in the last timestep that resulted in remnants with $V_{\rm max}$ divided by the number of halos with the same $V_{\rm max}$ in the current simulation output, $R_{-}\equiv N_{m-}(V_{\rm max})/N_{h}(V_{\rm max})$.  That is $R_-$ measures the fraction of halos that are a remnant of a merger that occurred in the last Gyr. The observational equivalent of this definition is a merger rate measured using morphologically identified merger remnants.  It is also the appropriate rate to compare with potential post-merger populations such as starbursts and luminous AGN.   A forward looking specific merger rate, $R_+$, is defined as the number of halos in the $V_{\rm max}$ range of interest that will experience a major merger in the next timestep divided by the number of halos in the same $V_{\rm max}$ range in the current timestep, $R_{+}\equiv N_{m+}(V_{\rm max})/N_{h}(V_{\rm max})$.  The forward looking merger rate measures the fraction of halos that will become the MMP of a major merger in the next Gyr.  The observational equivalent of $R_{+}$ is  a merger rate measured using pair counts. In these definitions $N_{m-}$ and $N_{m+}$ distinguish mergers that happened in the previous Gyr from those that will happen in the next Gyr. 

\section{Results} \label{results}
In this section results are presented for several different measurements of the evolution of and environmental dependence of the major merger rate.  The physical motivation for exploring each of these measurements is presented in this section, but a full discussion  of each result is postponed until \S\ref{discussion_m}. In what follows, several different classes of dark matter halos are discussed.  A `distinct' halo is any halo that is not a subhalo.   A `subhalo' is bound to and resides within the virial radius of a distinct host halo. A distinct halo may host subhalos, and therefore be a `host' halo, but it cannot itself be a subhalo.  While $V_{\rm{max}}$ may be used to designate the maximum circular velocity of any halo, $V_h$ is used specifically for either distinct or host halos and $V_s$ specifically for subhalos.

\subsection{Evolution of the Merger Rate} \label{evolution of the merger rate}

\begin{figure}
\begin{center}
\includegraphics[width = 6in]{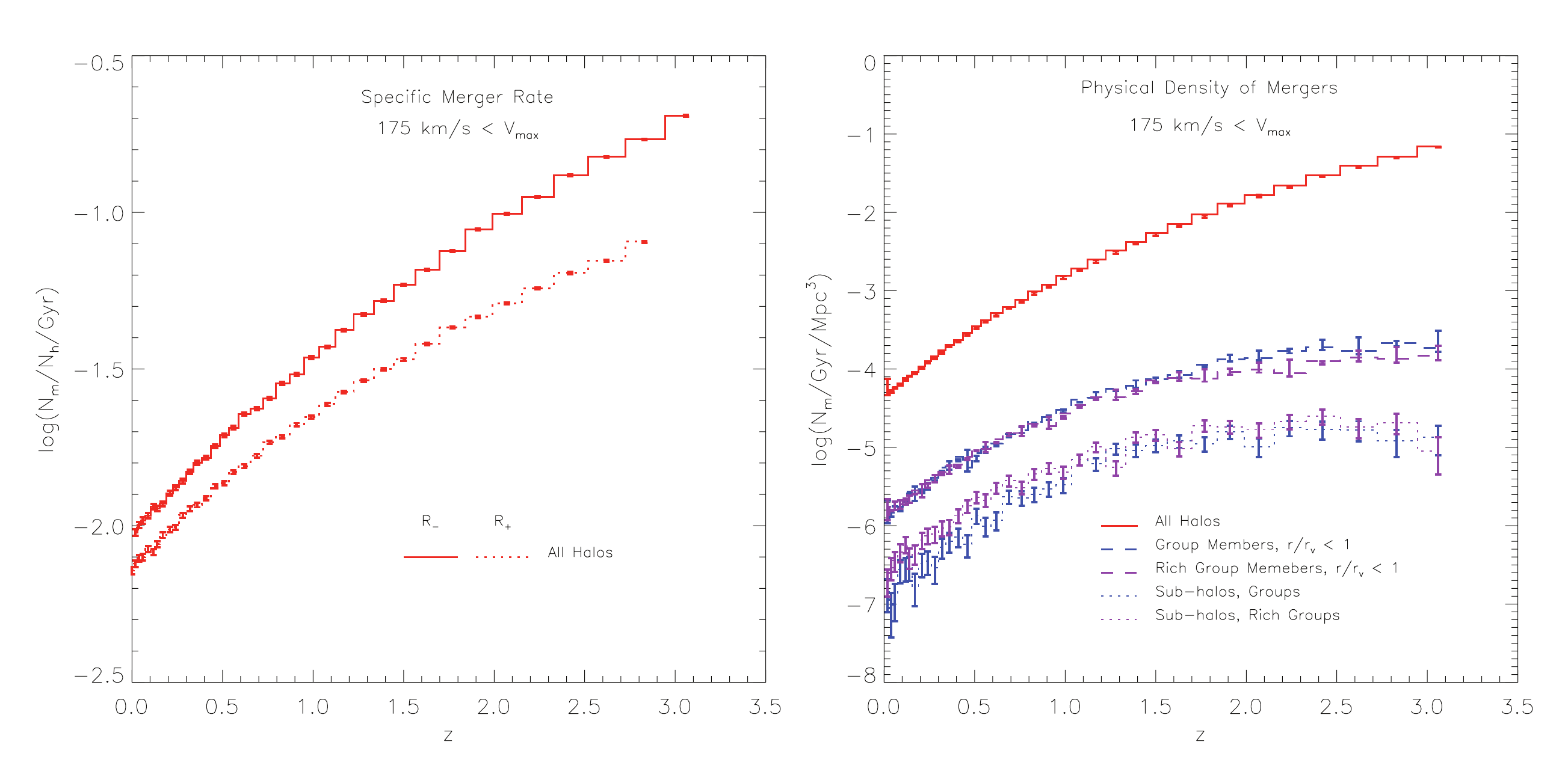}
\caption{Evolution of the merger rate.  The left hand panel displays the evolution of the specific merger rate, the merger rate per halo, for all halos. The forward looking rate, $R_+$, and the backward looking rate, $R_-$ are both plotted, where $R_+$ is the fraction of all halos with $V_{\rm max}>175\rm{km/s}$ that will become the most massive progenitor of a major merger in the next Gyr and $R_-$  the fraction of all halos with $V_{\rm max}>175\rm{km/s}$ that are a remnant of a major merger that occurred in the last Gyr.  These two definitions correspond to merger rates determined using counts of close pairs and morphologically identified merger remnants respectively.  The right hand panel displays the physical number density of major mergers per Gyr  resulting in remnants with $V_{\rm max}>175\rm{km/s}$.  The red solid line corresponds to all halos.  The blue and purple lines correspond to mergers occurring in `groups' and `rich groups' (See the text for group halo definitions).  The solid lines include both mergers between two subhalos and mergers between a subhalo and the host halo.  The dashed lines include only mergers between two subhalos.}\label{evolution}
\end{center}
\end{figure}

Figure~\ref{evolution} presents the evolution of the merger rate for all halos with $V_{\rm{max}}>175\rm{km/s}$.  The right panel displays the evolution of both specific merger rates, $R_+$ and $R_-$.  Comparing $R_+$ and $R_-$ demonstrates the differences one might expect when comparing the evolution of the merger rate as determined using different observational techniques, such as identifying merger remnants morphologically and immanent mergers using pair counts.   The evolution of the forward looking rate has a shallower slope than the backward looking rate. The backward merger rate, $R_{m-}\propto(1+z)^{2.2\pm0.05}$ and the forward merger rate, $R_{m+}\propto(1+z)^{1.82\pm0.01}$.  Due to completion issues at high z the actual slopes may be somewhat steeper.  This is discussed in \S\ref{completion}.  The difference between the slopes of the two measures of the merger rate, $R_-$ and $R_+$, is to first order set by the evolution of the  $V_{\rm{max}}$ function.  As is shown below, at fixed redshift the merger rate is relatively independent of $V_{\rm{max}}$.  In addition, the merger rate and number of halos change only slightly between consecutive simulation outputs. It follows that
\begin{displaymath}
\frac{R_-}{R_+} \approx \frac{R_+(t-dt)}{R_+(t)} \frac{N_h(V_{\rm{MMP}}, t-dt)}{N_h(V_{\rm{rem}}, t)} \approx \frac{N_h(V_{\rm{MMP}})}{N_h(V_{\rm{rem}})}
\end{displaymath}
where $V_{\rm{MMP}}$ is the typical $V_{\rm{max}}$ for the most massive progenitor of merger remnants with $V_{\rm{rem}}$.  This ratio is always greater than one, $V_{\rm{MMP}}<V_{\rm{rem}}$.  It increases with redshift as the $V_{\rm{max}}$ function shifts towards lower values of $V_{\rm max}$ and the local slope of the $V_{\rm{max}}$ function becomes steeper.  This is consistent with the relationship between $R_-$ and $R_+$ observed in Figure~\ref{evolution}.

The left panel of Figure~\ref{evolution} shows the evolution of the physical density of mergers.  The physical number density of merger remnants with $V_{\rm{max}}>175\rm{km/s}$ increases with redshift like $n_m\propto(1+z)^{5.44\pm0.02}$.  For comparison the physical number density of halos with $V_{\rm{max}}>175\rm{km/s}$ evolves as  $n_h\propto(1+z)^{3.13\pm0.03}$.  Both of these results differ from their mass selected counterparts.  In the case of a mass defined halo sample, the evolution of the halo number density above a lower mass limit is shallower than $(1+z)^3$ as halos are built up over time and the mass function evolves towards higher masses.  Similarly, the peak in the physical density of mergers at intermediate or high redshift seen in mass selected halo samples is not observed for a $V_{\rm{max}}$ selected halo sample.  The decline in the density of mass selected mergers at high z is due to the increasing scarcity of halos above a given mass as z increases.  The differences between these two types of samples can be reconciled.  The relationship between $V_{\rm{max}}$ and halo mass is evolving.  As the redshift increases $M_v(V_{\rm{max}})$ decreases.  The halo density and merger rate of lower mass halos are being measured at high z.
 
\subsection{Mergers Between Subhalos and Their Host Halos}\label{S sub host}

\begin{figure}
\begin{center}
\includegraphics[width = 6in]{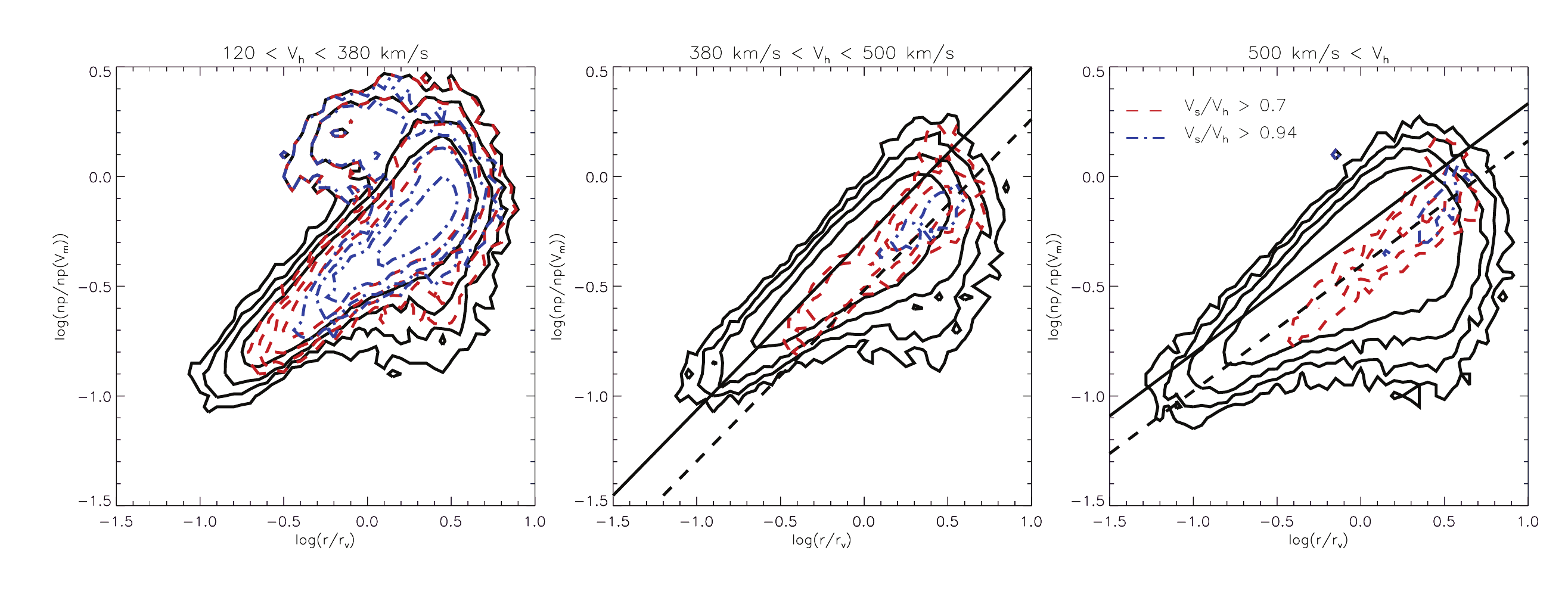}
\caption{Tidal stripping of subhalos at z=0.  Logarithmically spaced contours of subhalo number density in the $\log[np/np(V_{\rm max})]$ versus $\log(r_o/r_{vh})$ plane, where $np/np(V_{\rm max})$ is a well defined surrogate for $m_{ts}/m_{vs}$, as discussed in the text, and $r_o/r_{vh}$ is the subhalo's distance from the center of the host halo in units of the host halo's virial radius. In the three different panels, contours are shown for subhalos residing in hosts with $V_h$ values corresponding to galaxy-like host halos, group-like host halos, and rich group and cluster-like host halos.  The black contours are for all subhalos with $V_s>120\rm{km/s}$.  The colored contours are for subhalos whose merger with the host would be counted as a major merger.  Contours are shown for $V_s/V_h > 0.7$ and $V_s/V_h>0.94$, roughly corresponding to pre-tidal stripping mass ratios of 1:3 and 1:1.2.  With the exception of some subhalos  with high $V_s/V_h$, all subhalos are similarly tidally stripped.  The thin diagonal lines show the predicted relationships between $m_{ts}/m_{vs}$ and $r_p/r_{vh}$ for NFW halos for group-like (center panel) and cluster-like (right panel) hosts from~\citet{Mamon00}, where $r_p$ is the peri-center of the subhalo's orbit.  The right-hand line has been shifted to higher $r_o/r_{vh}$ by 0.3.} \label{tidal contours}
\end{center}
\end{figure}

\begin{figure}
\begin{center}
\includegraphics[width = 6in]{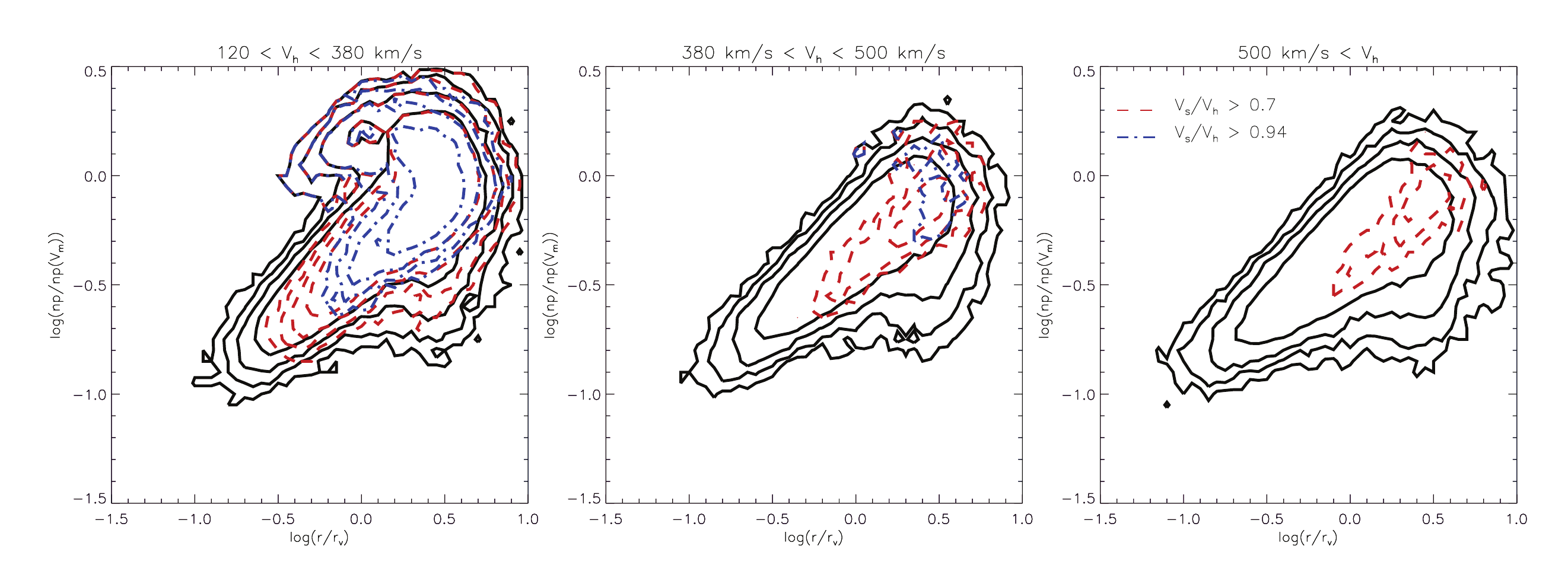}
\caption{Tidal stripping of subhalos at z=1.  The same as Figure~\ref{tidal contours}, but for $z=1$.}\label{tidal contours z1}
\end{center}
\end{figure}

When a halo is accreted by a more massive host halo and becomes a subhalo, it is potentially subject to tidal stripping, dynamical friction, unbound and bound collisions with other subhalos, and may eventually merge with the host halo.  The results presented in the next two subsections examine these issues in the light of their influence on the major merger rate.  Figures~\ref{tidal contours} and~\ref{tidal contours z1} begin by demonstrating that tidal stripping of subhalos is occurring in the Millennium Simulation. These figures show subhalo number density contours in the  $\log[np/np(V_{\rm{max}})]$ versus $r_o/r_{vh}$ plane at $z=0$ and $z=1$, where $r_0/r_{vh}$ is the current orbital radius of the subhalo divided by the virial radius of the host halo.  Contours are shown for all subhalos, and for subhalos with $V_s/V_h>$ 0.7 and 0.94, corresponding to approximate, pre-accretion, mass ratios, $m_{vs}/M_{vh}$, of 3:1 and nearly 1:1 (1:1.2), where $m_{vs}$ is the virial mass of the subhalo and $M_{vh}$ the virial mass of the host halo. Here $\log[np/np(V_{\rm{max}})]$ is a measure of the tidal stripping a subhalo has experienced.  Specifically, $np$ is the number of particles that remain bound to the subhalo and $np(V_{\rm{max}})$ is the typical number of bound particles in a distinct halo with the same $V_{\rm{max}}$.  Distinct halos exhibit a well defined correlation between $np$ and $V_{\rm{max}}$ with small but measurable scatter, as expected. Therefore,  $\log[np/np(V_{\rm{max}})]$ is a well defined substitute for $\log(m_{ts}/m_{vs})$, where $m_{ts}$ is tidal mass of the subhalo.  The three panels correspond to subhalos residing in galaxy-like host halos, $175 < V_h(\rm{km/s})<380$, group-like host halos, $380 < V_h(\rm{km/s})<500$, and a combination of rich group-like host halos, $500 < V_h(\rm{km/s})<950$ and cluster-like host halos, $950 < V_h(\rm{km/s})$.  Later `group-like' and `cluster-like' host halos will be distinguished from halos that actually host groups and clusters.

The least massive progenitors of sub-host major mergers are subject to the same tidal stripping as all subhalos.  In Figures~\ref{tidal contours} and~\ref{tidal contours z1} it can be clearly seen that subhalos with $V_s/V_h$ values that for distinct halos would corresponding to 3:1 and lower mass ratios are accreted and effectively tidally stripped by the larger, host, halo. The exception to this is the case of subhalos with $V_s/V_h$ values near 1:1 which show a spur at high $\log[np/np(V_{\rm{max}})]$ within the virial radius of the host.  The density contours for subhalos with $V_s/V_h$ corresponding to major sub-host mergers lie above the densest contours for all halos, consistent with these subhalos as a small random subset of all halos.  While these subhalos are found primarily at large radii, there is no significant difference in the distribution of $V_s/V_h$ at different $r_o/r_{vh}$. 

\begin{figure}
\begin{center}
\includegraphics[width = 6in]{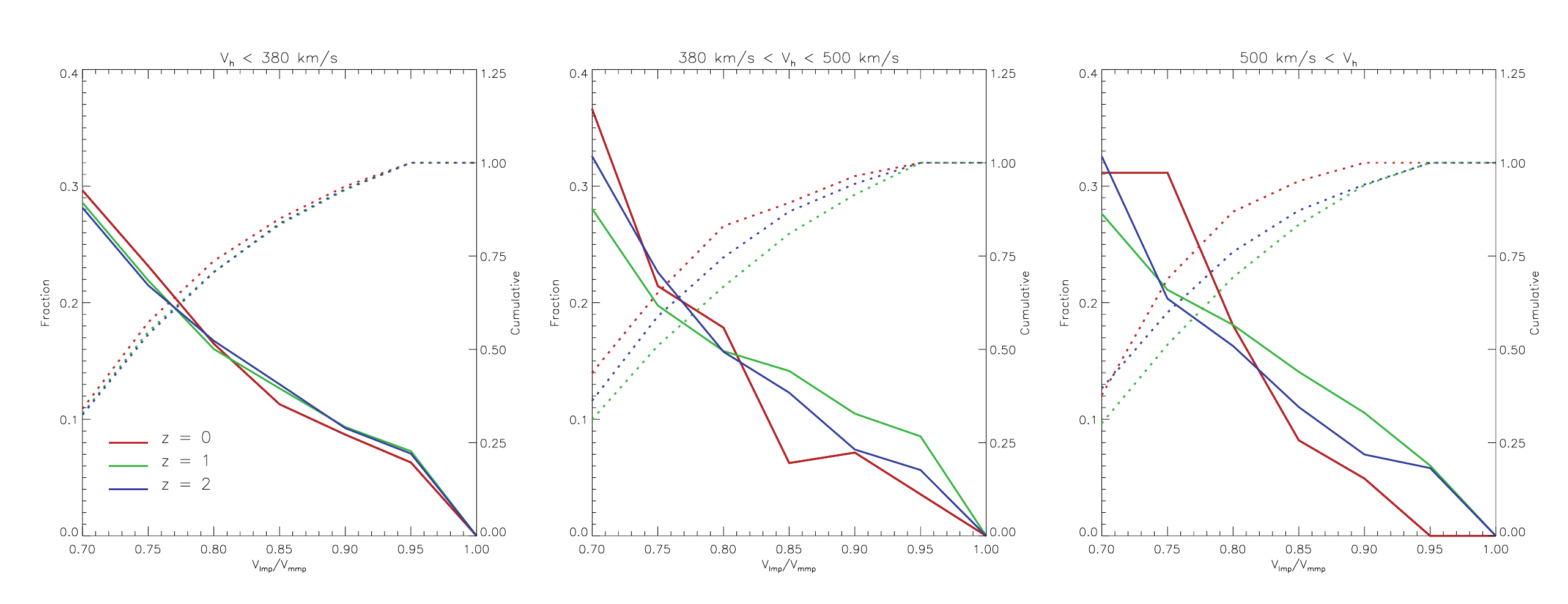}
\caption{Distribution of $V_{\rm LMP}/V_{\rm MMP}$ for major mergers at $z=0$, 1, and 2.   The three panels show the distribution split by merger remnant $V_{\rm max}$.  The same ranges are used as in Figures~\ref{tidal contours} and~\ref{tidal contours z1}.  The solid lines are histograms of the fraction of major mergers in each $V_{\rm LMP}/V_{\rm MMP}$ bin and the dashed lines are the cumulative probability distribution of $V_{\rm LMP}/V_{\rm MMP}$ values.  Mergers are heavily dominated by lower values of $V_{\rm LMP}/V_{\rm MMP}$, and it is safe to assume that the vast majority of major mergers are true sub-host mergers in which the subhalo merger partner has been tidally stripped before completely merging with the host halo.}\label{ratios}
\end{center}
\end{figure}

Figure~\ref{ratios} shows the distribution of $V_{\rm{LMP}}/V_{\rm{MMP}}$ for all major mergers resulting in galaxy-like, group-like, and rich group and cluster-like remnants.  Major mergers are dominated by the $V_{\rm max}$ equivalent of mass defined 2:1 and 3:1 mergers.  Only a few percent of mergers are close to 1:1 mergers.  Comparing Figure~\ref{ratios} with Figures~\ref{tidal contours} and~\ref{tidal contours z1} demonstrates that most mergers occur in the regime where the least massive progenitor (LMP) is first accreted as a subhalo, and tidally stripped, before it merges with the MMP, its host halo.  It is assumed in what follows that the vast majority of mergers can be cleanly separated into sub-host and sub-sub mergers.   

\begin{figure}
\begin{center}
\includegraphics[width = 6in]{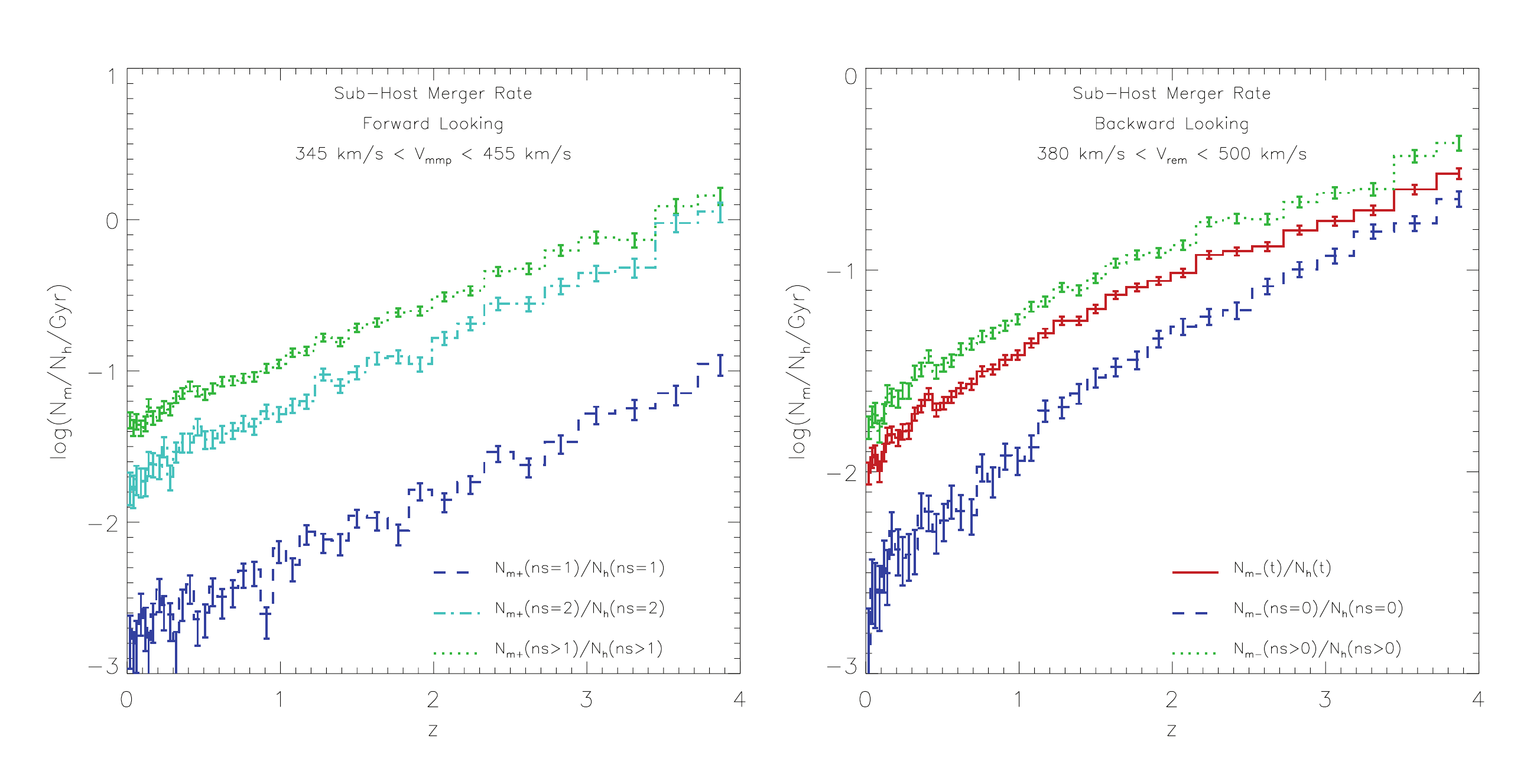}
\caption{The effect of interactions between subhalos on the major merger rate.  The left panel shows $R_+$ for sub-host mergers in hosts with $345<V_h\rm{(km/s)}<455$ split by the number of subhalos in the host.  The dark blue long-dashed line shows $R_+$ for mergers between a host and a lone subhalo.  The light blue dot-dashed line shows $R_+$ for sub-host mergers of hosts with two subhalos, and the green dotted line shows $R_+$ for sub-host mergers of hosts with at least two subhalos.  Introducing a second subhalo increases $R_+$ by a factor of $\approx 10$, rather than $\approx 2$, as would be expected if interactions between subhalos had no effect on the merger rate.  
The right panel demonstrates how this is reflected in the relationship between the number of subhalos and the frequency of merger remnants for halos with group-like $V_{\rm max}$. The dark blue dashed line shows the merger remnant frequency for halos with no subhalos with $V_s>120\rm{km/s}$.  These remnants would be the result of mergers between the host and a single bright subhalo.  The green line shows the frequency of merger remnants among group-like halos with at least one bright subhalo and the red line shows the frequency among all group-like halos. Merger remnants should be more likely than average galaxies to have a least one fainter close companion and should show enhanced correlations on small scales when cross correlated with a fainter galaxy sample.}\label{Rm ns}
\end{center}
\end{figure}

Figure~\ref{Rm ns} presents both the forward looking, left panel, and backward looking, right panel, merger rate for sub-host mergers resulting in group-like halos.  Dynamical friction is not effective at degrading the orbits of  tidally stripped subhalos~\citep{Boylan07}. Despite this the $R_+$ for sub-host mergers can reach values as high as a merger every one to a few Gyr, as can be seen in the left panel of Figure~\ref{Rm ns}.  This suggests that the dynamics of these systems are not completely captured by considering a combination tidal stripping and dynamical friction alone.  In addition to interacting with their host, subhalos also interact with each other, motivating us to examine whether the presence of multiple subhalos affects the sub-host merger rate.  

In the left panel of Figure~\ref{Rm ns}, $R_+$ is plotted for host halos with $345<V_{\rm{max}}(\rm{km/s})<455$ and 1, 2, and greater than 1 subhalo, $ns=1$, 2, and $>1$.  The subhalo counts include all subhalos with $V_{\rm{max}}>120\rm{km/s}$.  To compute this rate, at each time-step the number of major merger remnants in the next time-step with a most massive progenitor with $345 <V_{\rm{max}}(\rm{km/s}) < 455$ and with one fewer subhalo than the $ns$ of interest, is divided by the number of halos in the current time-step with $345 < V_{\rm{max}}(\rm{km/s})< 455$  and the desired $ns$.  This assumes, for example, that all merger remnants with no subhalos are the result of a merger between a host and its single subhalo.  The time-step is small enough that the occurrence of a sub-host merger and the accretion of a new subhalo within the same time-step should be rare enough not to effect these results.  When this does occur it will bias against the result observed.  The right panel of Figure~\ref{Rm ns} shows the number of merger remnants with $380 < V_{\rm{max}}(\rm{km/s})< 500$ divided by the number of halos in this $V_{\rm{max}}$ range in the same timestep.  This $V_{\rm{max}}$ range roughly corresponds to the remnants of the mergers plotted in the left-hand panel.  In this panel the number of merger remnants with the $ns$ of interest is divided by the number of all halos with the same $ns$.  While the left hand panel best captures any dynamics, the right hand panel is easiest to compare with observations.  Particularly in groups, it is a backward looking rate, a rate calculated using mergers that have already happened, that is easiest to observe.  A forward looking merger rate necessarily relies on pair counts.  

The specific forward looking merger rate in host halos that have two subhalos  is approximately 10 times the specific merger rate in host halos with only one subhalo pre-merger. The distribution of $V_s/V_h$ for host halos within this narrow range in $V_h$ does not depend on $ns$; a host halo with two subhalos is only twice as likely to host a subhalo with $V_s>0.7V_h$, a potential major merger LMP.  If the existence of multiple subhalos did not affect the dynamics of sub-host mergers, then host halos with two subhalos would experience major mergers twice as often. This effect is reflected in the backwards looking merger rate. The frequency of merger remnants among host halos with $ns>0$ is several times larger than the frequency among host halos without a subhalo with $V_{\rm{max}}>120\rm{km/s}$.  Conversely, at low redshifts, where the effects of subhalo completion are less important, a second subhalo is present for 80-90\% of the sub-host mergers in group-like halos. Differences between the two panels of Figure~\ref{Rm ns} are mainly due to the ratios $N_h(ns=1)/N_h(ns=0)$ and $N_h(ns=2)/N_h(ns=1)$.  In the right panel, mergers of hosts with lone subhalos are normalized by the number of halos with no subhalos rather than the number of  halos with one subhalo. These ratios evolve with redshift.  Interactions between subhalos are examined in the next section and a physical mechanism for enhancing the sub-host merger rate is advanced.

\subsection{Mergers Between Subhalos}\label{subs}

Observationally, galaxies can be separated into `isolated' galaxies and members of groups or clusters.  Previous works based on mass defined subhalo samples find that the specific merger rate is enhanced in groups.  By selecting subhalos based on mass, these works treat tidal stripping by excluding tidally stripped subhalos.  In this section, we take the next step by using a $V_{\rm{max}}$ defined subhalo sample to determine if the observed specific merger rate is indeed expected to be higher in groups than for isolated galaxies.

Observational group finding algorithms seek to identify bound virialized objects by finding over-densities in the projected galaxy density within a narrow redshift range. The simulation equivalent is a host halo with its subhalos. Accordingly, groups in the simulation are defined as group-like halos, $380 < V_h(\rm{km/s}) < 500$, that host at least two subhalos with $V_{\rm{max}} > 120\rm{km/s}$ within their virial radius.  Group members include the host halos and its subhalos.  Rich groups are defined as rich group-like halos, $500 < V_h(\rm{km/s}) < 950$, with at least two subhalos. The observational equivalent to these simulation groups would be a catalog of groups containing at least three member galaxies, the central galaxy and two satellites.  

In addition to the evolution of the physical number density of all mergers, the right panel of  Figure~\ref{evolution} displays the physical number density of mergers of all group members. Figure~\ref{evolution} also includes the physical number density of mergers between two subhalos,  or sub-sub mergers.  Specifically, Figure~\ref{evolution} shows the number of merger remnants that are subhalos of group hosts and of rich group hosts.  Mergers within groups account for a small percentage of all mergers. Sub-sub mergers make up roughly 10\% of the mergers that occur in group environments and a fraction of a percent of all mergers.  The simulation groups host 2-3 subhalos on average, depending on the redshift, and the specific merger rate for subhalos is 20-30 times lower than the specific merger rate for the host halos of groups.  This is in direct contrast with the results of studies using mass selected subhalos. In host halos with $V_h<320\rm{km/s}$, it is somewhat more common for a subhalo to be a merger remnant.  This occurs most frequently for subhalos with $V_s/V_h\approx1$ which does not correspond to a true sub-sub merger, but rather a sub-host merger in disguise.  Not only are mergers between subhalos rare in an absolute sense, the specific merger rate of subhalos, or the frequency of merger remnants among subhalos, is low.  Major mergers are dominated by sub-host mergers.

\begin{figure}
\begin{center}
\includegraphics[width = 6in]{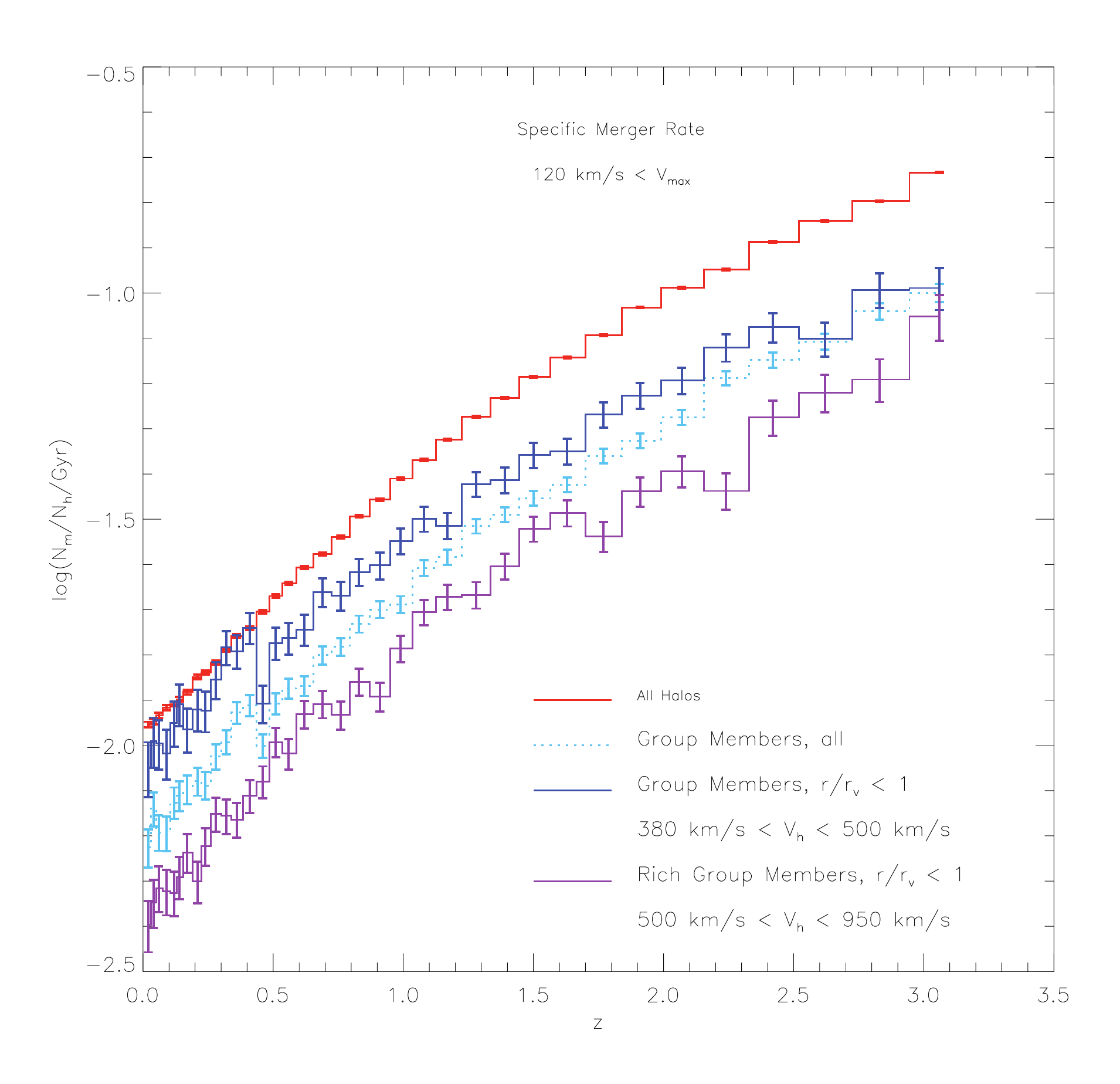}
 \caption{The evolution of the specific major merger rate, $R_-$, for  `group' and `rich group' members is compared to that for all halos.  Groups and rich groups are defined in the text.  `Members' includes both the central host halo and its subhalos. Contrary to previous expectations, the specific merger rate in groups is lower than in the field. }\label{groups}
\end{center}
\end{figure}

Together the sub-host merger rate and sub-sub merger rate determine the specific merger rate among group members.  In Figure~\ref{groups}, $R_-$ is plotted versus redshift for all halos,  group members (`Group Members, $r/r_v<1$'), and members of  rich groups ( `Rich Group Members, $r/r_v<1$').    An alternate definition of a group is also used that requires the host halo have two `subhalos' within the dark matter FOF group surrounding the central host halos and includes all subhalos within the FOF group as group members.  Groups thus defined are designated `Group Members, all'.  In a similar fashion, observational group finding algorithms vary in how conservatively they define groups and group members.  As can be seen in Figure~\ref{groups}, the specific merger rate among group members is in fact lower than for all, or for `isolated', halos in a $V_{\rm{max}}$ selected halo sample.  It is also likely to be suppressed in groups constructed from luminosity selected galaxy samples.  The specific merger rate is lower for rich group members than group members and the specific rate for generously defined groups, `Group Members, all', is lower than for conservatively defined group members, `Group Members, $r/r_v<1$'.

Previous analytical work has shown that for mass selected subhalos the sub-sub merger rate decreases drastically with host halo mass while increasing with subhalo mass.  For example \citet{Mamon00} finds $R_+ \propto n G^2 m_s^2/V^3_h$. These generic dependences arise for a range of specific analytical cross sections. It is interesting to check whether similar scalings occur for $V_{\rm max}$ selected subhalos in order to determine if one should expect to observe these trends for real galaxies.  Somewhat surprisingly, no such trend is observed for the $V_{\rm max}$ selected subhalos.  The forward looking merger rate, $R_+$, for sub-sub mergers is significantly suppressed relative to the sub-host merger rate and shows no dependence on either $V_h$ or $V_s$.  The lower specific merger rate for group members of rich groups, as opposed to the subhalos in rich groups, is due to the increased contribution of subhalos.  

The absence of a trend with $V_h$ is somewhat startling as the relative velocities of subhalos are higher in clusters than in groups.  The key to this puzzle may be the relatively large masses of the subhalos studied.  The analytic result assumes that the number of subhalos scales with $M_{vh}$, resulting in a constant number density of subhalos.  In the simulation, while the average number of subhalos with $V_{\rm{max}}>175\rm{km/s}$  does increase approximately as $V_h^3$, the average number of these somewhat massive subhalos ranges from $<1$ to a few.  For group-like host halos there are multiple hosts that host only one such subhalo for every host with two or more such halos.  The number of halos that host multiple large subhalos and are therefore capable of hosting a sub-sub merger increases with $V_h$, off setting the increase in relative velocities between subhalos.  A simulation with a finer mass resolution may reveal a correlation between $R_+$ and $V_h$.  It is less surprising that no trend is observed between $R_+$ for sub-sub mergers and $V_s$.  First, while sub-sub mergers may be more likely among subhalos with higher post-tidal stripping masses, and subhalo mass and $V_s$ are loosely correlated, there is significant scatter in the mass-$V_s$ relation of subhalos. Second, major mergers are defined differently in the two scenarios.  For $V_s$ defined subhalos, the case of two merging subhalos with similar $V_s$ is marked as a major merger even if one subhalo is newly arrived and the other significantly tidally stripped.  Conversely, a merger of two subhalos with similar masses and different degrees of stripping will not be considered a major merger if the $V_s$ are significantly different. The sub-sub merger rate may depend on the subhalo mass without such a dependence being observed in a $V_s$ defined subhalo sample. As a final consideration, the relatively low number of mergers between subhalos results in poor statistics, and a weak potential dependence on $V_h$ or $V_s$ would therefore not be observed.  This leaves open the possibility that a weak trend observed in a large survey would not be inconsistent with the simulation results. 

\subsection{Merger Rate vs Local Halo Density}\label{S local density}

\begin{figure}
\begin{center}
\includegraphics[width = 6in]{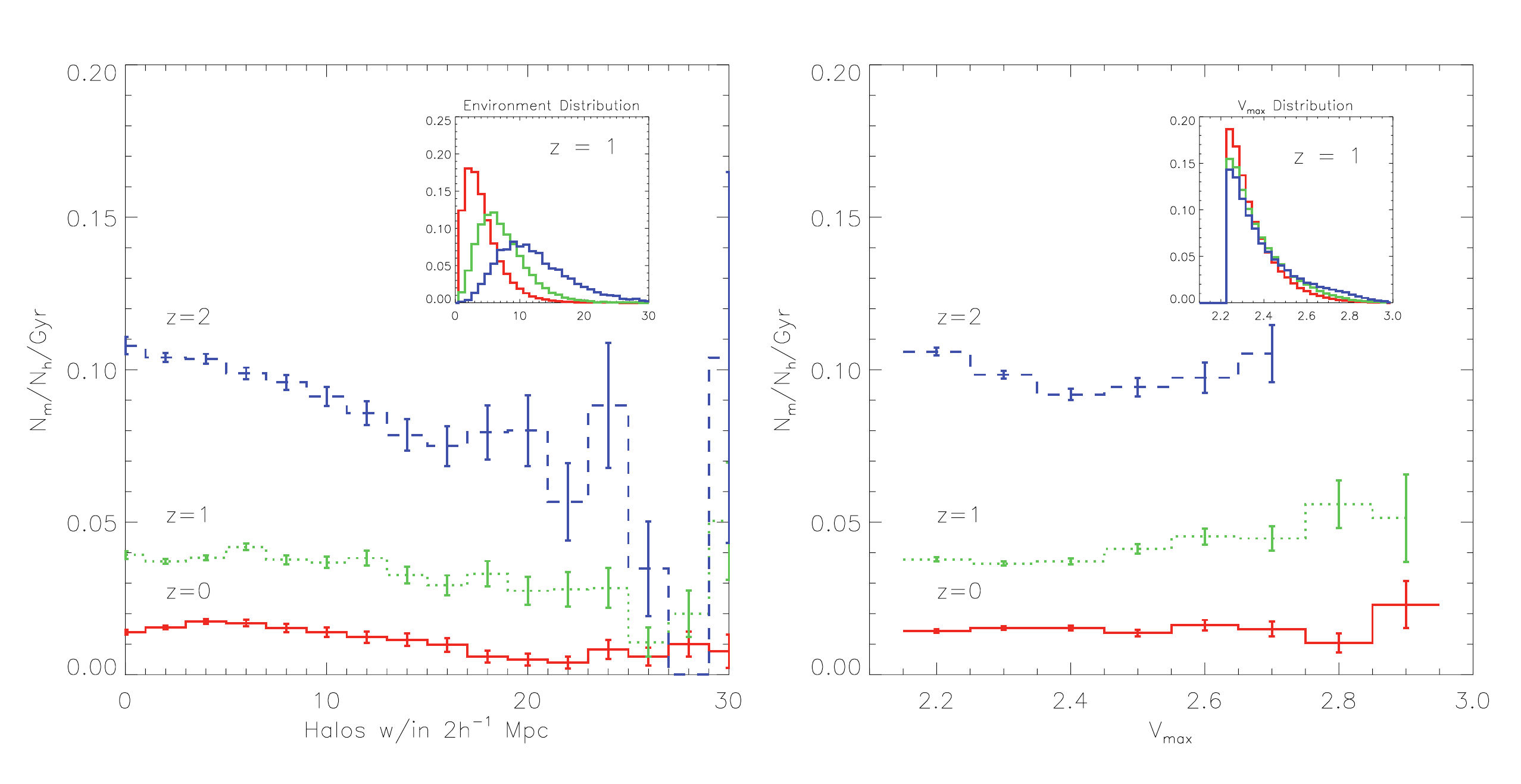}
\caption{Merger rate versus local halo number density and $V_{\rm max}$.  The left panel displays $R_-$ versus the local halo number density, measured by counting all halos with $V_{\rm max}>120\rm{km/s}$ within a $2h^{-1}\rm{Mpc}$ comoving sphere, for all halos with $V_{\rm max}>175\rm{km/s}$ for $z=0$, 1, and 2.  The right panel shows $R_-$ versus $V_{\rm max}$ for halos in all environments.  Insets are for orientation.  The left inset shows the distribution of local densities for $z=1$ for three $V_{\rm max}$ ranges, galaxy-like ($175 < V_{\rm{max}}(\rm{km/s})<380$), group-like ($380<V_{\rm{max}}(\rm{km/s})<500$), and rich group-like ($500 < V_{\rm{max}}(\rm{km/s}) < 950$).  This inset clearly illustrates halo biasing.  The right inset shows the distribution of $V_{\rm max}$ in three environment ranges, 0-4, 5-9, and 10-14 halos within $2h^{-1}\rm{Mpc}$ comoving, displaying the counterpart of halo biasing. The merger rate is independent of $V_{\rm max}$ and, for environments typical of most halos, of environment.}\label{env}
\end{center}
\end{figure}

Figure~\ref{env} shows $R_-$ versus local density for all halos with $V_{\rm{max}}>175\rm{km/s}$ in the left panel, and in the right panel displays $R_-$ versus $V_{\rm{max}}$ for all halos in all environments.  Both are shown for $z=0$, 1, and 2.   The local density is the count of all halos, both distinct and sub, with $V_{\rm{max}}>120\rm{km/s}$ within a sphere of radius $2h^{-1}\rm{Mpc}$ comoving. The halo sample is complete to lower values of $V_{\rm{max}}$ than the merger sample, simply because both progenitor halos must be resolved in order to identify a merger.  This allows a the use of a lower $V_{\rm{max}}$ to measure the environment, which has the advantage of sampling the environment more accurately. The insets in Figure~\ref{env} are provided for orientation. 

A correlation between the merger rate and either local environment or $V_{\rm{max}}$ would be straight forward to compare with observations.  Unfortunately, there is little dependence of the merger rate on environment or $V_{\rm{max}}$ independently. The merger rate does decline slightly above local densities of 15-20 at $z=0$ and 1, and at slightly lower densities at $z=2$. As can be seen by referring to the insets, however, the vast majority of halos reside in environments with lower densities, where the merger rate is constant with local density.

\begin{figure}
\begin{center}
\includegraphics[width = 6in]{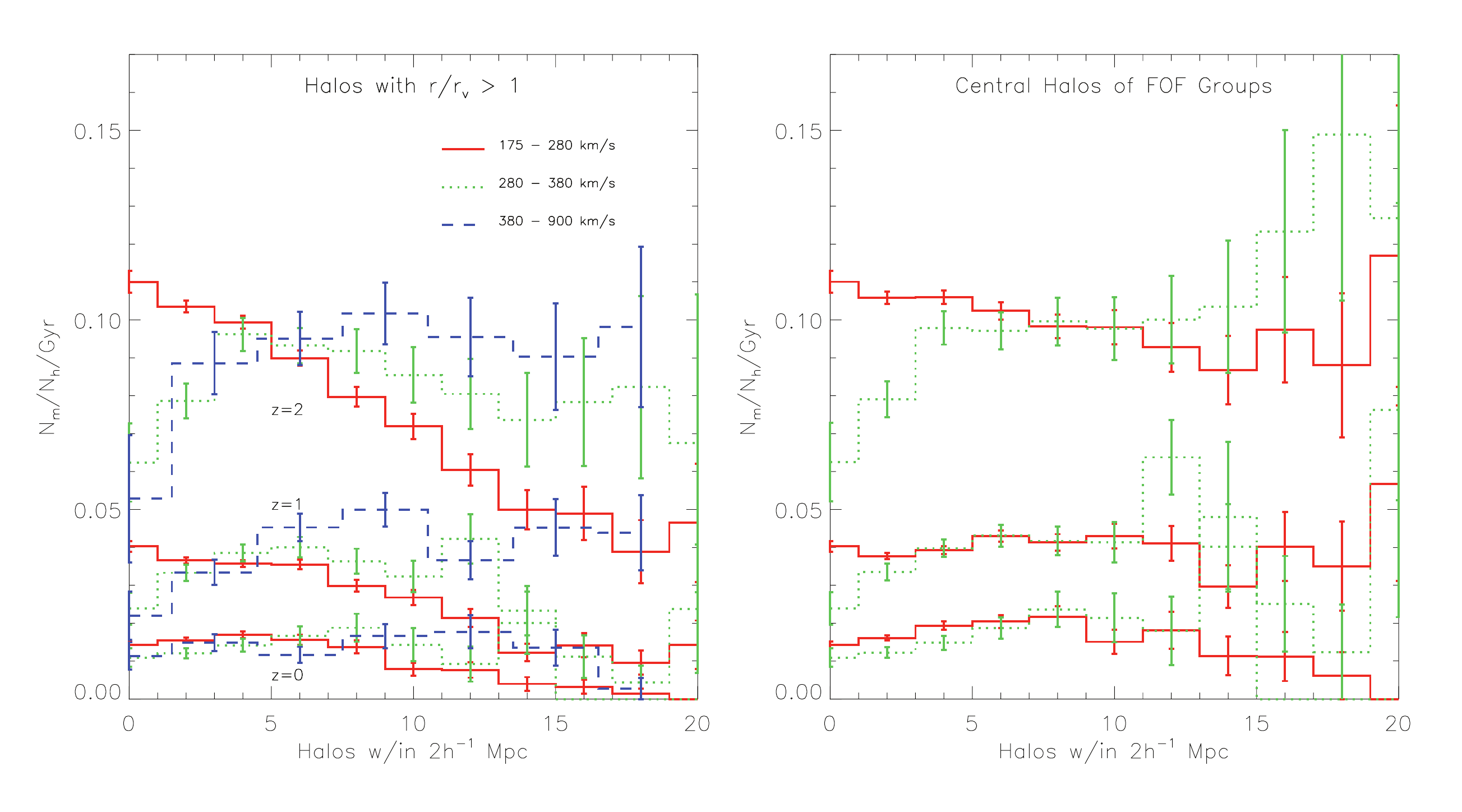}
\caption{Merger rate versus local halo number density for halos grouped by $V_{\rm max}$. In both panels, $R_-$ is plotted versus the local density at $z=0$, 1, and 2 for halos in three $V_{\rm max}$ ranges. The middle $V_{\rm max}$ range, the green dotted line, evolves from moderately biased at $z=2$ to unbiased at $z=0$. The left panel includes both central halos and non-central halos that are beyond the virial radius of the central halo of their dark matter FOF group.  In this panel, the major merger rate shows an assembly-type bias with $R_-$ decreasing with local density for halos with low $V_{\rm max}$ and increasing with local density for halos with high $V_{\rm max}$.  
The right panel includes only halos that are the central halo of their dark matter FOF group, excluding `distinct' halos on the outskirts of more massive halos.  For halos with low $V_{\rm max}$, $R_-$ is observed to decrease with local density in the right panel, but not in the left.  The left panel is sensitive to the effects of a massive host halo on nearby distinct halos while the right panel is sensitive to the effect of the local environment on halos that dominate their immediate surroundings. Comparing the two panels provides insight into the possible physical mechanisms responsible for the major merger assembly bias}\label{bias}
\end{center}
\end{figure}

Recent work has shown that \textit{codependent} correlations between halo properties such as concentration or age and both $V_{\rm max}$, or mass, and local environment do exist in large N-body simulations~\citep{Wechsler06, Croton07, Gao07}. In Figure~\ref{bias} $R_-$ is plotted versus the local environment for halos in three $V_{\rm{max}}$ ranges for $z=0$, 1, and 2.  The $V_{\rm{max}}$ ranges used are $175 < V_{\rm{max}}(\rm{km/s}) < 280$, $280 <V_{\rm{max}}(\rm{km/s}) < 380$, and $380<V_{\rm{max}}(\rm{km/s})<900$.  These are different than the $V_{\rm{max}}$ ranges used previously and have been selected to highlight the evolution of the trends observed in Figure~\ref{bias}.  Halos in the intermediate $V_{\rm{max}}$ range evolve from moderately biased at $z=2$ to relatively unbiased at $z=0$ while halos in the upper and lower $V_{\rm{max}}$ ranges are biased and unbiased, respectively, at all three redshifts.  The left panel includes all halos that are either the central halo of their dark matter FOF group or are beyond the virial radius of their central halo. These halos are selected because linear theory predicts that, when true subhalos are excluded, $R_-$ at fixed $V_{\rm{max}}$ should be independent of local environment.  In contrast, in the simulation the halos with the lowest values of $V_{\rm{max}}$ show a slight trend of decreasing $R_-$ with local density, and the halos with the highest values of $V_{\rm{max}}$ show a slight trend of increasing $R_-$ with local density.  

Distinct halos in the vicinity of a more massive halo, that is halos under the dynamical influence of a nearby massive halo, may be subject to different dynamics than more isolated distinct halos.  The right panel of Figure~\ref{bias} shows $R_-$ versus the local density for only halos that are the central halo of their FOF group, hence excluding distinct halos that are on the outskirts of a more massive halo. Only the two lower $V_{\rm max}$ ranges are plotted in the right panel.  The right panel of Figure~\ref{bias} emphasizes the interaction between central halos, which dominate their immediate environment, and environment on intermediate scales. There are two points of note when comparing the two panels of Figure~\ref{bias}.  First, for the lowest $V_{\rm{max}}$ range $R_-$ versus local density is significantly flatter in the right panel.  Second, in the right panel the trend for the mid $V_{\rm{max}}$ range evolves from a $R_-$ that increases with local density at $z=2$ to an absence of a trend at $z=0$.  This is in contrast to a relatively flat relation between $R_-$ and environment at all redshifts for the same $V_{\rm{max}}$ range in the left panel.  

For Figures~\ref{env}  and~\ref{bias}, it is appropriate to include both distinct halos and subhalos in the local environment definition. Ideally, one would plot the trends shown in Figure~\ref{bias} for dark matter over-density rather than halo number density, which unfortunately has no observational equivalent.  This ideal is best approached by including subhalos as the local halo count is thereby implicitly weighted by both the mass and proximity of nearby halos.  A massive halo near the edge of the $2h^{-1}\rm{Mpc}$ sphere contributes to the local density by as many of it's subhalos as lie within the sphere.  As was observed in Section~\ref{S sub host}, however, the sub-host merger rate increases with the number of subhalos so it is prudent to examine how this correlation affects the results presented here.  

The correlation between the number of subhalos and the merger rate may bias the results in the sense that increasing the number of subhalos will increase both the merger rate and the local halo count. This biases against observing the trend observed for halos in the low $V_{\rm{max}}$ range, but has the same sense as the trend observed for the high $V_{\rm{max}}$ range.  The number of subhalos in a host is typically smaller than the range of environments plotted, which is expected given that the virial radii of these halos is smaller than $2h^{-1}\rm{Mpc}$.  While checking for this bias we found a direct correlation between the number of subhalos in a host and the local environment excluding the hosts own subhalos. This is consistent with the results of \citet{Wechsler06}.  This correlation may drive the observed trend between the major merger rate and the local environment given the significant increase in the major merger rate for hosts with multiple subhalos.

\section{Discussion}\label{discussion_m}

This section discusses possible physical mechanisms behind the results presented in the previous section and compares the results to observations from the literature.

\subsection{Evolution of the Merger Rate}

Figure~\ref{evolution} emphasizes that for distinct halos it is the growth of structure, set by cosmology and the initial matter distribution, that sets the merger rate.  The evolution of both $R_-$ and $R_+$ is significantly flatter than that of $n_{h}$.  While $n_h\propto(1+z)^{3.13\pm0.03}$,  $R_- \propto (1+z)^{2.2\pm0.05}$ and $R_+ \propto (1+z)^{1.82\pm0.01}$.  The evolution of the major merger rate likely depends on the sample selection, is affected by subhalo completion, and is complicated by any evolution in the $L$-$V_{\rm max}$ relation.  We therefore refrain from predicting a slope of the evolution of the merger rate, but these caveats aside expect that the major merger rate evolves more slowly than the halo number density.  Additionally, within a single galaxy sample, the merger rate as measured using close pair counts and using morphologically identified mergers will evolve differently. When the possible effects of sample selection are also taken into account, comparing the evolution of the merger rate as determined using different techniques on different samples should be undertaken with great caution.

\citet{GKK} consider a third measure of the evolution of the specific merger rate in which only halos that belong to a MMP chain of a halo identified at $z=0$ are considered. They select halos using $V_{\rm max}$ at $z=0$. They find that this rate evolves as $(1+z)^{\sim3}$, which is steeper than our result for all major mergers.  One might expect that the specific merger rate along the MMP chain and the specific merger rate for all halos would evolve with the same slope.  This assumes that the likelihood of a halo remaining on the MMP chain is independent of merger history.  That is, that both major merger remnants and all halos are each as likely as the other either to continue as a MMP or to become an LMP. That~\citet{GKK} measure a steeper rate than found in this work suggests that halos which undergo major mergers are more likely to survive as part of a MMP chain.  This dependence is only suggested by these results.  While the merger rate measured here includes halos with the same $V_{\rm{max}}$ at all redshifts, the average $V_{\rm{max}}$ of the progenitor halos in a MMP chain decreases with redshift. This comparison emphasizes the importance of both measuring the merger rate as~\cite{GKK} have in order to compare morphologies at $z=0$ to merger history and also measuring the merger rate as is done here in order to compare to the observed evolution of the merger rate.  It is not correct either to integrate the merger rate observed here in order to compare with morphologies at $z=0$ or to compare observed merger rates to the evolution measured by~\cite{GKK}.

Previous observations of the evolution of the major merger rate, using both pair counts and morphologies, have found a wide range of slopes when the evolution is characterized as $R_m\propto(1+z)^m$.  Most are flatter than $m=3$, with some that are actually consistent with no evolution.  For both methods, previous results can be roughly split between groups that find $m=2-4$~\citep{LeFevre00, Patton02, Conselice03, Cassata05, Kampczyk07} and $m<1$~\citep{Carlberg00, Lin04, Lotz08}. There are many differences between these studies including sample selection, merger selection, pair selection, and whether the evolution is measured within a single sample or by comparing two or more samples.  In general, stricter dynamical requirements on pair counts result in steeper evolution.  \citet{Berrier06} find that pair counts in simulations show little evolution because the decline in the merger rate is counterbalanced by the build up of groups and clusters.  Measurements at high redshift rely on morphologically identified mergers which can be complicated by resolution, loss of low surface brightness features, and by observing galaxies at non-ideal rest frame wavelengths.  Automated morphological measurements may also have difficultly separating major mergers from galaxies with asymmetric star formation.  While individual groups do attempt to quantify and correct for these effects, the disparity between different measurements indicates there is more work to be done.  Our prediction that the major merger rate evolves with a flatter slope than the number density of galaxies cannot yet be compared to such disparate observations.

\subsection{Groups and Clusters}

In the MS subhalos of all $V_s/V_h$ are tidally stripped of similar mass fractions, which is consistent with theoretical expectations.  A simple model of tidal stripping, which assumes that subhalo orbits are nearly circular and that the host-halo and the newly accreted subhalos share a self similar density profile, for example an NFW profile, predicts that $m_{ts}/m_{vs}=M_h(r_o)/M_{vh}$, where $m$ denotes the mass of the subhalo before and after tidal stripping, $M$ the mass of the host, and $r_o$ the orbital radius of the subhalo within the host.  \citet{Mamon00} presents an analytical estimate of the tidal mass of a subhalo with an NFW profile orbiting in an NFW host in which the correlation between mass and concentration and the typical orbits of subhalos in NFW hosts have been considered (their equation 8).  In Figure~\ref{tidal contours} the diagonal lines in the left and right panels correspond to this analytical estimate in groups and clusters respectively.  The left line is simply the estimate given in~\citet{Mamon00}, and the right line is this estimate shifted to higher $\log(r_o/r_{vh})$ by 0.3.   As long as some subhalos have made one peri-centric passage, the average peri-centric radius, $r_p/r_{vh}$ of subhalos found at each $r_o/r_{vh}$ is lower than $r_o/r_{vh}$ itself. Scatter about this relationship can be introduced by scatter in the relationship between $V_{\rm max}$ and $np$, deviations from circular orbits, and scatter in the concentrations of the host-halo and subhalo density profiles.  Subhalos that lie in the bulge to the lower left of the contour plot are likely subhalos on radial orbits that have passed through their peri-center at least once. The contours presented in Figures~\ref{tidal contours} and~\ref{tidal contours z1} are clearly consistent with representing tidal stripping of subhalos in the MS, that is the MS appears to be correctly capturing tidal stripping of subhalos.  

The MS also appears to correctly capture the interplay between tidal stripping and dynamical friction. Before a major merger can occur, the stripped LMP must lose its orbital energy and angular momentum.  The dynamical friction time scales of these stripped subhalos, which retain only 10-30\% of their original mass, are quite long.  \citet{Boylan07} consider the interplay between tidal stripping and dynamical friction by running high resolution simulations of a lone subhalo that is accreted by and eventually merges with its host. Their results indicate that the merger time-scale for \textit{major mergers} likely ranges from $\approx3.5$ to 7 Gyr, with longer time-scales for lower $m_{vs}/M_{vh}$ and higher circularities.  Major mergers are dominated by mergers with initial mass ratios closer to 3:1 than 1:1, and are therefore expected to have the extremely long merger time-scales closer to the long end of this range.   Our results using the MS are consistent with those of~\cite{Boylan07}.  Assuming that all host halos with a single subhalo at $z=1$ whose single subhalo has $V_s/V_h>0.7$ experience a merger at $z=0$ correctly predicts the major merger rate among halos with a single subhalo at $z=0$.  This is of course a very rough consistency check, but it indicates that the MS is correctly capturing both tidal stripping and dynamical friction.  

The significantly enhanced sub-host major merger rate when all hosts, including those with multiple subhalos, are considered indicates that some dynamics beyond tidal striping plus dynamical friction is at work.  This is confirmed by considering the effect on the sub-host merger rate of introducing a second subhalo; doubling the number of subhalos increases the sub-host merger rate by a factor of 10, not a factor of 2.  Mergers, or bound collisions, between subhalos are rare, likely due to the small merger cross section of tidally stripped halos.  Unbound collisions between subhalos in groups and clusters are likely quite common, and an unbound collision between subhalos with opposing angular momentum has the potential to reduce the orbital energy and angular momentum of both subhalos.  Assuming elastic unbound collisions, angular momentum can be effectively scattered between subhalos, but not effectively cancelled. An analytical treatment of this problem that assumes inelastic collisions finds no enhancement for the sub-host merger rate from sub-sub interactions~\citep{Penarrubia05}.  In reality, however, sub-sub interactions are extremely inelastic, as indicated by studying the mass loss induced by these interactions~\citep{Knebe06}.  In the case of inelastic collisions, subhalos with opposing angular momenta can effectively dump both orbital energy and angular momentum when they experience an unbound collision.  Given inelastic collisions between subhalos, the presence of another, or several, other subhalos of comparable mass or $V_{\rm{max}}$ may considerably shorten the time it takes a subhalo to merge with its host.   That this is occurring requires further study, but it is both physically plausible and consistent with our results.  Correctly predicting the sub-host merger rate requires considering tidal stripping, dynamical friction, and interactions between sub-halos.

While previous results have suggested that the specific merger rate may be enhanced in groups, for a $V_{\rm max}$ selected halo sample the merger rate in the simulation equivalent of groups is suppressed relative to that for distinct halos.   The sub-host merger rate is enhanced in groups, but subhalos dominate numerically in groups and have extremely low specific merger rates. Note that the specific merger rate of group `members', as opposed to subhalos, declines with group mass, reflecting the increasing numeric dominance of the subhalos. When attempting to compare an observed specific merger rate in groups with the rate in the simulations it should be kept in mind that the predicted rate may depend quite heavily on the construction of the group catalog.  As can be seen in Figure~\ref{groups}, the specific merger rate is lower for generously defined groups than for conservatively defined groups.

Based on our results we can make several observational predictions about the major merger rate. First, we find that mergers are dominated by sub-host mergers and that the specific merger rate of sub-host mergers is 20-30 times larger than for sub-sub mergers.  As a result, the major merger rate among all group members is suppressed relative to the field, contrary to previous predictions.  Second, the strong enhancement of the merger rate in halos hosting multiple subhalos can be tested observationally. The simulation results predict that merger remnants should be more likely than an average galaxy, hosted in a halo of the same mass as the merger remnant, to have a fainter near-by companion, which is likely bound to the merger remnant.  Similarly, when observationally identified mergers are cross correlated with a fainter galaxy sample, mergers are predicted to show a significantly enhanced correlation on small scales, that is scales corresponding to the one-halo term of the correlation function. 

\citet{McIntosh07} study the major merger rate in groups by identifying morphologically disturbed close pairs of galaxies.  They find that a slight majority of mergers involve the central galaxy of a group and that the specific merger rate for central galaxies is $\approx3\%$ and for satellite galaxies is less than $1\%$.  Their observations agree qualitatively, but not quantitatively, with the simulation results presented here.  When translating the simulation results into observational predictions, any scatter between the $V_{\rm max}$ of the host halo and luminosity, as well as any uncertainties in the observations, will tend to result in observational results that are less extreme than the simulation results.  That is, while the difference between the specific merger rate of host halos and of sub-halos may differ by a factor or 20-30 in the simulations, a factor of 10 or less difference in the observations may be consistent. \citet{McIntosh07} also find that the sub-sub specific merger rate declines with the mass of the host group.  Their groups have an average of $\approx5$ members and all have multiple sub-halos.  They therefore avoid the contribution of group-like halos with only one sub-halo and appear to capture the effect of increasing relative sub-halo velocities with group mass.  

Previous theoretical and observational results have indicated that starbursts and AGN show an excess of companion galaxies on small scales. \citet{Thacker06} study the AGN-galaxy cross correlation in a SPH simulation in which AGN are fueled by major mergers and find an excess at small scales compared to the galaxy-galaxy correlation.  \citet{Goto05} cross correlate 266 E+A galaxies in the SDSS with the SDSS imaging catalog and find that E+A galaxies have an excess of companions on scales $<150\rm{kpc}$ when compared to normal galaxies.  \citet{Serber06} preform a similar analysis for AGN and find an excess of companions on similar scales.  Previous studies of companion frequencies have suggested that tidal interactions with the companion trigger activity.  We offer an alternative scenario in which activity was triggered by a major merger which was facilitated by the companion.  The companion galaxies must also have lost considerable angular momentum, hence their small distances from the active galaxy.

\subsection{Major Merger Rate and the Assembly Bias}

While the major merger rate shows no independent correlation with either local environment or $V_{\rm{max}}$ for the $V_{\rm{max}}$ range considered,  $R_-$ does correlate with the local environment at fixed $V_{\rm max}$.  The sense and size of this correlation is similar to so called `assembly biases' seen for related halo properties such as concentration and age.  The results presented in \S\ref{S local density} provide insight into the potential physical mechanisms driving this correlation.

There is no theoretical reason that the major merger rate should not depend on environment or on $V_{\rm max}$, and it may be a coincidence that no such correlations are observed.  Halos with lower $V_{\rm max}$ values are found on average in lower density environments and the average  $V_{\rm max}$ increases with local overdensity.  If this were not the case, then the observed codependent correlations would require that the merger rate decline with increasing local density.  These trends appear to conspire to produce no observed dependence on environment when the full $V_{\rm max}$ range is included, a result which may not hold to lower $V_{\rm max}$.  A slight correlation between the merger rate and the halo mass has been observed previously in the MS~\citep{Fakhouri08}.  The absence of this trend in the results presented here is likely due to the scatter between halo mass and $V_{\rm{max}}$. 

In linear theory the formation history of a halo of fixed mass, which is captured by $R_-$, does not depend on environment.  This is not what is observed in the \textit{Millennium Simulation}. Similar deviations from the expectations of linear theory have recently been observed in simulations for other halo properties such as halo formation time, concentration, number of subhalos, subhalo mass function,  and halo angular momentum~\citep{Wechsler06, Croton07, Gao07}.  The correlation between the likelihood of having experienced a recent major merger and the local environment may be yet another facet of what has been termed the `assembly bias'.  These correlations are all closely related through halo's assembly histories.  Younger halos are also less concentrated~\citep{NFW, Wechsler02, Zhao03}, and a halo that has recently experienced a major merger is younger.  Halos that host a higher than average number of subhalos are more likely to experience, and to have experienced, a major merger.  Previous studies have all found that the size of the assembly bias is small, and the results here are consistent with this.  For the range of local environments in which the majority of halos find themselves, the merger rate is relatively independent of environment. 

Linear theory is based on the spherical collapse model, in which the radial evolution of shells of dark matter is determined by the mass contained within each shell. Tidal forces are neglected, as are the orbits and interactions of previously virialized halos within these shells.  The role of tidal forces has been previously acknowledged; the halo mass function can be predicted correctly only if tidal forces are included in an average sense~\citep{Sheth99}.  Tidal forces, halo orbits, and interactions between bound halos, all analogs of the dynamics within host halos, must contribute to the non-linear trends observed in larger simulations. 

Tidal stripping within FOF groups is occurring beyond the virial radii of the central halos.  Figures~\ref{tidal contours} and~\ref{tidal contours z1} include all FOF non-central halos.  The relationship between tidal mass and $r_o/r_{vh}$ extends beyond the viral radius of the FOF central halo to radii of a few $r_{vc}$.  Note that this is predicted by the analytical estimate of~\citet{Mamon00}.  When a halo gets within a few virial radii of a more massive halo, the weak tidal forces in the outskirts likely suppresses the major merger rate of the halo by decreasing accretion onto them. In the right panel of Figure~\ref{bias}, which excludes halos on the outskirts of more massive halos, no correlation between major merger rate and environment is observed at $z=0$ and 1 for the low $V_{\rm max}$ halos and the correlation is considerably reduced at $z=2$.
 
Halos which are destined to be accreted by a larger halo have always been bound to the larger halo in that their orbital energy with respect to the host halo has always been negative.  These halos are not on radial orbits, however.  Major mergers in simulations have a range of initial orbital parameters, with a distribution of orbital circularities, $j/j_c(E)$, that peaks near 0.5~\citep{Khochfar06}.  These bound halos are subject to dynamical friction and to unbound collisions with other halos that may cause them to be prematurely, when compared to linear theory, accreted by the central host halo. Despite the fact that these halos are only mildly tidally stripped, dynamical friction is likely ineffective in the low dark matter densities beyond the virial radius of the central halo.  Both relative velocities between orbiting halos and the number density of bound objects are also lower beyond the virial radius.  Unbound collisions between these objects, which will be rarer but more effective, may be an effective means of driving halos into the central host.  If this is the case, then group-like halos in dense environments will experience somewhat enhanced accretion rates of bound objects, have larger subhalos counts, and higher major merger rates.  This of course assumes that the group-like halo is not itself in the vicinity of a yet more massive halo.

For central halos, an increase in the local density corresponds to an increase in the number of halos between one and a few $r_v$ that are under the influence of the central halo. In contrast, for non-central `distinct' halos, increasing environment corresponds to increasing average tidal forces. The high $V_{\rm max}$ halos are all central halos, but the middle $V_{\rm max}$ range includes both central and non-central `distinct' halos.  This is also true for the low $V_{\rm max}$ halos, but for these halos $2h^{-1}\rm{Mpc}$ is considerably more than a few $r_v$.   At $z=2$ when the central halos in the middle $V_{\rm max}$ range are isolated, the merger rate increases with local environment.  Poor statistics prevent this trend from being measured with clear significance, but the difference between the panels in Figure~\ref{bias} is suggestive.  Such a difference is not observed for lower redshifts, perhaps because it is increasingly more likely for a group-like halo to be in the vicinity of a cluster-like halo even if it is a central halo.

This scenario may also provide insight into the evolution of the major merger assembly bias.  The following discussion applies more generally to any trend that is driven by the proximity of a more massive halo. At any redshift, the youngest, most massive, biased halos universally dominate their local dynamics, corresponding to $R_-$ increasing with local halo number density. For smaller halos the chances of being in the vicinity of one of these halos increases with environment.  While in less dense environments $R_-$ for these smaller halos may still increase with environment, in rich environments it will decline.  As time advances and these larger halos become less biased, the smaller halos will find themselves in the vicinity of a more massive halo in less extreme environments, resulting in an $R_-$ that decreases with environment in all environments.  As the growth of structure advances further, halos with increasingly higher values of $V_{\rm max}$ form, and the halos that were once universally dominant find themselves transitioning into a regime in which $R_-$ declines with environment.  In this picture, the likelihood of showing a galaxy-like or a group-like correlation between $R_-$ and environment correlates with the bias of the halos.  This discussion is meant only to consider the issues involved and suggest an aspect of this picture that may be rewarding to explore further. One general characteristic of this scheme is that the $V_{\rm{max}}$ at which the dependence of $R_-$ on environment transitions from group-like to galaxy-like evolves towards higher values with time, which is typical of the assembly biases. 

Observing the trends between $R_-$ and local halo density may be possible either directly or through clustering. It can be predicted that, when cross correlated with a fainter galaxy sample, galaxy populations corresponding to the brightest merger remnants will show enhanced clustering on scales larger than the typical virial radius of the remnants.  Cross correlating with a fainter galaxy sample allows the environment on scales of a few virial radii to be adequately sampled.  Confronting previously identified assembly biases observationally is difficult both because the correlation is small and because the relevant halo properties may only loosely correlate with galaxy properties.  One exception, the number of subhalos a group hosts, has its own difficulties due to constraints imposed by estimating group masses~\citep{Berlind06b}.  The major merger rate may therefore be an interesting way to directly probe the assembly biases as a class.  

The picture presented here may also contribute to a physical understanding of any assembly type bias that is observed.  For example,~\citet{Berlind06b} find that groups whose central galaxies are bluer than average, or rather less red than average, are biased on large scales compared to average groups.  While bluer galaxies are typically considered `younger', blue colors suggest the central galaxy has recently received either gas or young stars.  This does not correlate in a straight forward way with any of the previously observed assembly biases.  In contrast, we find not only that major merger remnants are biased, but that for merger remnants in dense environments the average time between subhalo accretion and a sub-host merger is reduced.  Therefore, sub-host mergers in these environments are more likely to bring young stars, and possibly gas, to the central galaxy.  

\subsection{Subhalo Completion} \label{completion}

In order to determine when our results may be impacted by the artificial disruption of subhalos, we apply the criteria of \citet{Klypin99}, which were discussed in \S\ref{description of simulation}, to the relevant subhalos.  We begin by determining whether our results are limited by mass resolution or force/spatial resolution.  For the MS, a tidal radius of two times the spatial resolution corresponds to $10h^{-1}\rm{kpc}$.   Assuming, for simplicity, a static subhalo density profile, mass resolution is the limiting factor for subhalos.  That is, for the particle mass and spatial resolution of the MS, subhalos drop below the 30 particle limit before their tidal radius reaches $10h^{-1}\rm{kpc}$.  A typical halo with $V_{\rm max}=100\rm{km/s}$, the lowest relevant $V_{\rm max}$, that has been stripped to the 30 particle limit has a tidal radius significantly larger than $10h^{-1}\rm{kpc}$, so this conclusion likely holds in the realistic case that stripping alters the subhalo density profile. To check this conclusion we follow subhalos in the bound particle number versus orbital radius plane and find the redshift at which subhalos with $V_{\rm max}=120\rm{km/s}$, the lower limit of our catalog,  begin dropping below the 30 particle limit.   We do not appear to be loosing any subhalos before this point.  For each of the minimum $V_{\rm max}$ values considered below, the same result applies; the limiting condition is that a typical subhalo retain at least 30 bound particles.  When relevant subhalos begin dropping below the 30 particle limit before reaching orbital radii at which a merger with the host is presumed imminent, subhalo retention has become an issue.

To estimate if and when subhalos with a given $V_{\rm max}$ drop below the 30 particle limit we use the analytical estimate of~\citet{Mamon00} for tidal stripping of NFW halos;
\begin{displaymath}
m_{ts} \approx m_{vs}a_r(r_o/r_{vh})^{b_r}
\end{displaymath}
where $m_{ts}/m_{vs}$ is the tidal mass of the subhalos as a fraction of the initial virial mass, $r_o/r_{vs}$ the orbital radius scaled by the virial radius of the host halo, and $a_r$ and $b_r$ are chosen to reflect departures of the halo density profile from self similarity.  As can be seen in Figure~\ref{tidal contours}, this provides a fair description of tidal stripping in the MS.  By using the correlation measured in the MS between $V_{\rm max}$ and bound particle number for distinct halos, we can estimate the $r_0/r_{vh}$ to which a typical halo with a given $V_{\rm max}$ retains 30 particles.  This relation evolves with redshift and subhalos of a given $V_{\rm max}$ are lost at increasing $r_0/r_{vh}$ as the redshift increases.  The pertinent $V_{\rm max}$ are 100 km/s for sub-sub mergers with remnant $V_{\rm max} > 175{\rm km/s}$, 120 km/s for subhalo counts, and 210 km/s for sub-host mergers in `group-like' halos with $V_{\rm max}>380\rm{km/s}$.   In the simulation, 90\% of the least massive progenitors of major mergers have $V_{\rm max}$ values greater than 0.6 times that of the merger remnant. Hence the values of  100 km/s and 210 km/s. Typical subhalos with $V_{\rm max} = 210\rm{km/s}$ survive numerical effects to $r_o/r_{vh}=0.1$ out to $z>4$.  Subhalos with $V_{\rm max}=120\rm{km/s}$ survive to $r_o/r_{vh}<0.1$ at $z=0$,  $r_o/r_{vh}\approx0.2$ at $z=1$, and $r_o/r_{vh}\approx0.4$ at $z=2$.  Typical subhalos with $V_{\rm max}=100\rm{km/s}$ survive to $r_o/r_{vh}\approx0.25$ at $z=0$ and $r_o/r_{vh}\approx0.5$ at $z=1$.

These estimates show that we are clearly pushing the ability to track subhalos in the MS.  Most of the important results, however, are quite robust against the effects of subhalo completion. The largest exception is the case of mergers between subhalos.  Subhalo mergers are more likely to happen before subhalos are significantly tidally stripped, which mitigates the effects of loosing potential least massive merger progenitors to artificial evaporation at moderate radii.  That said, the result that subhalo mergers are rare must rest solely on results from redshifts near $z=0$.  The most robust result is that the presence of multiple subhalos drastically reduces the time-scale for sub-host mergers.  It is the presence of bound structures in the simulation that drives the merger rate.  It is therefore appropriate to separate halos based on the presence or absence of such a structure, regardless of whether a subhalo would exist in the hypothetical case of better mass resolution.  The related result that an assembly-type bias is observed in the major merger rate is somewhat less robust as it depends on identifying the particular halos in which subhalos counts and the sub-host merger rate are enhanced.  The fidelity of this identification starts fading near $z\approx1$ as subhalos with $V_{\rm max}\approx120\rm{km/s}$ begin to be subject to artificial evaporation.  In the absence of a correlation between environment and the probability of a subhalo evaporating, however, this effect only introduces noise.  This may degrade the observed trend but cannot create it.  To the extent that environment and subhalo loss may be correlated, the initial bound particle counts of fresh subhalos are likely suppressed in over-dense environments, biasing against the observed result.  

The evolution of the merger rate for halos with $V_{\rm max}>175\rm{km/s}$ may be strongly affected by subhalo incompletion.  Completion affects both the measured evolution of $R_+$ and $R_-$ in all environments and the comparison between environments. In all environments, the first subhalos lost are not the least massive progenitor halos of the sub-host mergers, which dominate the merger rate.  Rather they are the lower $V_{\rm max}$ subhalos which facilitate the sub-host major mergers.  These halos are not only missed by the halo finder, they are truly absent having been artificially dissolved.  This results in a major merger rate that is increasingly underestimated at high redshifts; an effect that becomes important for galaxy-like halos at lower redshifts than group-like halos.  In galaxy-like halos a second effect comes into play when the least massive progenitors of sub-host mergers are artificially dissolved, resulting in a major merger occurring prematurely.  The combination of these effects may be quite complex,  yet in Figure~\ref{evolution}, the merger rate for all halos shows a smooth evolution.   Subhalos also play a role in group identification, complicating the comparison of the major merger rate between different environments.  As completion begins to affect subhalos with $V_{\rm max}\approx120\rm{km/s}$, the requirement that a `group' halo have at least two subhalos becomes increasingly stringent.  As some `real groups' are not identified, the physical density of mergers in groups is underestimated at high redshift and the measured slope of the evolution is too shallow.  The flattening of the physical number density of mergers in group seen in Figure~\ref{evolution} is likely in part due to completion.  The average number of subhalos in the groups also declines with redshift, which may be real, artificial, or a combination. Note that these effects refer only to the evolution of the merger rate in groups.  As long as the  specific sub-sub merger rate is significantly lower than the sub-host specific merger rate, the specific merger rate in groups will be suppressed compared to the merger rate for all halos.

\section{Conclusions}
We have made a detailed study of the environments of major mergers in the \textit{Millennium Simulation}~\citep{Springel05ms}.  Our goal in doing so is to provide a theoretical background for observational tests both to confront the dark matter simulations directly and to identify populations of merger remnants through their environments.  Comparing simulation results with observations correctly requires some caution.   The most robust observational tests rely on physically motivated comparisons between actual observations.  Such tests are likely to be remarkably insensitive to the details of the simulation.  Comparisons between observations that are motivated by the simulations, but which are not physically motivated may be interesting, but their interpretation will likely be difficult.  They may be useful for testing the accuracy of numerical simulations when subtle resolution effects are involved.  Finally, comparing simulation results to observations directly is tempting, but perilous.  When trying to make such a comparison extreme care must be taken to understand sample selection, both in the simulations and the observations, and to understand the effects of halo completion, with all subtleties included.  

We find that the evolution of the specific merger rate, both forward and backward-looking, is flatter than the evolution of the number density of halos.  This comparison can be made directly in the observations and has physical implications because it probes the growth of structure.  Comparing an observational result to the simulations also has the ability to probe the importance of subhalo completion on the simulated merger rate.  In the simulations subhalo completion affects the evolution of the major merger rate but not the evolution of the halo number density.  The discrepancy between the two slopes, $R$ and $n$, is large, however, and it is likely the difference cannot be accounted for by completion effects.  Comparing the slope of the evolution of the merger rate for galaxy samples with different luminosity limits may be another way to probe the effects of completion in the simulation as they become increasingly important for lower $V_{\rm max}$ halo samples.  In contrast, comparing the measured slope of the evolution of the merger rate from the simulations directly to an observed slope is unlikely to be rewarding as the simulated evolution is affected by completion, may depend on halo selection, and such a match would require modeling any evolution of the $V_{\rm max}$ - luminosity relation.  

Subhalos are quickly tidally stripped after they are accreted by their host unless $V_s/V_h\approx1$. One simple effect of this is that the cross section for sub-sub mergers is drastically reduced.  In the MS this is seen as a specific merger rate for sub-sub mergers that is 20-30 times lower than for sub-host mergers.  The specific merger rate among group `members' is therefore suppressed compared to isolated halos.  The mergers in groups are dominated by sub-host mergers, but the halo counts are dominated by subhalos.  Our first physically motivated prediction is that the specific major merger rate, that is the merger rate per galaxy, is suppressed in groups.  This contrasts with previous predictions based on mass selected sub-halos.  We caution that the exact specific merger rate measured in the simulated groups should not be compared directly to an observed rate because group selection can have a significant effect on the measured specific major merger rate.

Tidal stripping dramatically lengthens the time between accretion and the sub-host merger in host halos with a lone subhalo, which was seen both by~\citet{Boylan07} and in the extremely low specific major merger rates observed for single subhalo systems in the MS.  In halos that host multiple subhalos, the merger rate is enhanced by a factor of 10.  Subhalos experience inelastic collisions~\citep{Knebe06} during which subhalos with opposing angular momentum may be able to dump energy and angular momentum allowing subhalo interactions to facilitate sub-host mergers. Observing the effects of these interactions on the major merger rate is robust to subhalo completion issues as it is driven by the presence of bound objects in the simulation.  Merger remnants are significantly more likely than the average halo to host a subhalo.  The observational consequence of which is that the frequency of faint near-by companions should be enhanced for merger remnants. Similarly, when cross-correlated with a fainter galaxy sample, merger remnants should show an enhanced correlation on scales less than the virial radius of the remnants.  

The major merger rate was found to be uncorrelated with both local environment and $V_{\rm max}$ when each was studied independently. There is no physical motivation for this result, and it may not hold down to lower $V_{\rm max}$.   At fixed $V_{\rm max}$ however we found an assembly type bias in the backward looking merger rate.  For galaxy-like halos $R_-$ decreases with increasing density while for group-like halos it increases.  A physical mechanism for the major merger assembly bias was advanced in \S\ref{discussion_m}.  Tidal stripping extends beyond the virial radius of central halos and may be responsible for the decrease in the major merger rate of galaxy-like halos in dense environments.  In contrast, interactions between bound halos in group outskirts may enhance accretion onto group-like halos in dense environments, resulting in a positive correlation between subhalo counts and local environment~\citep[as seen by][]{Wechsler06}.  This correlation would drive an increase in the major merger rate for group-like halos in dense environments. For assembly biases in general,  the typical $V_{\rm max}$ or mass at which the cross over from galaxy-like to group-like behavior occurs increases with time, which is expected if they are driven by the likelihood of finding a more massive halo in the vicinity. We predict that major merger remnants that are the central galaxies of groups should show enhanced clustering on scales beyond the viral radius of the group.  Major mergers are a promising way of observing an assembly-type bias because the link between halo and galaxy properties is straight forward.

The physically motivated predictions presented above can be used to confront the simulation results with observations, provided that major merger remnants are conservatively identified.  We predict that merger remnants should show an excess of faint nearby companions, have enhanced clustering on small scales, and, for the central galaxies of groups, also have enhanced clustering on intermediate and large scales.  Using environment to test whether other populations are merger remnants will be complicated by the physics of these objects.  Three populations of interest are starbursts, K+A galaxies, and AGN, which are all considered possible stages of the evolution of a merger between two blue spiral galaxies into a red elliptical.  The correlation between dark matter mergers and these populations depends on the presence of substantial amounts of gas during the merger, which we do not attempt to track in the simulation.  In the case of mergers between two galaxy-like halos, one or both of the merging galaxies is likely to have gas.  For mergers involving a group-like halo, either merger progenitor may lack gas and the likelihood of the merger involving gas may depend on environment.  If the group-like halo hosts a hot inter-group medium, then the least massive progenitor, which generally spends a considerable time orbiting the group prior to merging, may be stripped of its gas.  Not all group-like halos host an IGM however.  The likelihood of a group hosting an IGM correlates with the morphology of the central galaxy, with blue central galaxies corresponding to a lack of an IGM and vice versa~\citep{Osmond04, Brough06}.  Early type groups, which host an IGM, may reside in different environments than the late type groups and hence the likelihood of a merger involving gas may depend on the local environment.  

Given the complicating gas physics, studying the frequency of fainter companions has clear advantages over clustering studies.  When studying small scales, i.e. the one-halo term, cross correlations weigh the contribution of each central galaxy by the number of subhalos, $ns$.  If the likelihood of being able to fuel a starburst or an AGN correlates with $ns$, then observing an enhanced clustering will depend on the sign and strength of this correlation.  When studying larger scales, the two-halo term, observing the predicted enhancement for group-like halos requires that gassy major mergers are a fair sample of all major mergers, which is likely not the case.  In contrast, studying the frequency of fainter companions does not require a fair sampling and weights each central halo equally.  

When studying the environments of AGN further physics comes into play.  AGN display a complex relationship between host mass or $V_{\rm max}$ and AGN luminosity which may depend on both environment and redshift~\citep{Hopkins06, Hopkins06b}.  The distribution of $V_{\rm max}$ for luminous AGN will be different than that of the underlying merger population if the AGN lifetime correlates with $V_{\rm max}$.  This will affect the clustering of these AGN, even in the absence of a correlation between environment and the merger rate.  This will also complicate constructing a comparison sample for studies of companion frequencies.  Determining whether AGN are mergers may require modeling AGN fueling and attempting to over constrain such models with several observational measures of environment, number density, and evolution.  Additionally, some AGN may be triggered by by alternate mechanisms, though merger driven AGN likely dominate at high redshift and bright luminosity~\citep[][and references therein]{Hopkins06c}.

Designing specific observational tests based on the results presented here is a topic for future work.  

\section{Acknowledgments}
This work was partially funded by  NSF grant 0707731. The authors would like to thank E. Scannapieco for useful discussions.  J.A.H. would like to thank S. Malhaotra, J. Rhoads, and the Arizona State University School of Earth and Space Exploration for their hospitality while this work was being completed.

The Millennium Simulation databases used in this paper and the web application providing online access to them were constructed as part of the activities of the German Astrophysical Virtual Observatory.

The particular structure of the database design which allows efficient querying for merger trees is described in: Lemson G. \& Springel V. 2006, Astronomical Data Analysis Software and Systems XV, ASP Conference Series, Vol. 351, C. Gabriel, C. Arviset, D. Ponz and E. Solano, eds.

\clearpage

\end{document}